\documentclass{IEEEtaes}
\usepackage{amsmath,amssymb,amsfonts}
\usepackage{url}
\usepackage{verbatim}
\usepackage{graphicx}
\usepackage{color,array,amsthm}
\usepackage{graphicx}
\usepackage{subfigure}
\usepackage{cite}
\usepackage{textcomp}
\usepackage{xcolor}
\usepackage{gensymb}
\usepackage{makecell}
\usepackage{stfloats}
\usepackage{multirow}
\usepackage{bm}
\usepackage{cuted}
\usepackage{algorithm}
\usepackage{mathrsfs}
\usepackage{algpseudocode}

\begin{document}

\title{Multipath Identification and Mitigation with FDA-MIMO Radar}

\author{Yizhen~Jia}
\author{Jie~Cheng}
\author{Wen-Qin Wang}
\member{Senior member, IEEE}
\author{Hui~Chen}
\member{Member, IEEE}

\receiveddate{Manuscript received September 22, 2023; revised XXXXX 00, 0000; accepted XXXXX 00, 0000.\\
	 \itshape (Corresponding author: Wen-Qin Wang)}

\authoraddress{Yizhen~Jia, Jie~Cheng, Wen-Qin Wang, and Hui~Chen are with School of Information and Communication Engineering, University of Electronic
	Science and Technology of China, Chengdu, 611731, P. R. China. Email: (jiayizhen@std.uestc.edu.cn, 	jiecheng567@gmail.com,
	wqwang@uestc.edu.cn, huichen0929@uestc.edu.cn,
	)
} 

\supplementary{Color versions of one or more of the figures in this article are available online at \href{http://ieeexplore.ieee.org}{http://ieeexplore.ieee.org}.}

\maketitle

\begin{abstract}
	In smart city development, the automatic detection of structures and vehicles within urban or suburban areas via array radar (airborne or vehicle platforms) becomes crucial. However, the inescapable multipath effect adversely affects the radar's capability to detect and track targets. Frequency Diversity Array (FDA)-MIMO radar offers innovative solutions in mitigating multipath due to its frequency flexibility and waveform diversity traits amongst array elements. Hence, utilizing FDA-MIMO radar, this research proposes a multipath discrimination and suppression strategy to augment target detection and suppress false alarms. The primary advancement is the transformation of conventional multipath suppression into a multipath recognition issue, thereby enabling multipath components from single-frame echo data to be separated without prior knowledge. By offsetting the distance steering vectors of different objects to be detected, the accurate spectral information corresponding to the current distance unit can be extracted during spatial spectrum estimation. The direct and multipath components are differentiated depending on whether the transmitting and receiving angles match. Additionally, to mitigate high-order multipath, the echo intensity of multipath components is reduced via joint optimization of array transmit weighting and frequency increment. The numerical results show that the proposed algorithm can identify multipath at different distances in both single-target and multi-target scenarios, which is superior to the general MIMO radar.
\end{abstract}


\section{INTRODUCTION}
A{\scshape s} smart cities evolve, the need for vehicular or unmanned aerial vehicle (UAV) array radar to automatically monitor buildings or vehicles within urban areas becomes increasingly critical. It is widely used in fields such as autonomous driving\cite{Chen2023milestones,AbdallahMultipath2021}, and outdoor positioning and tracking\cite{Luo2022antenna,Rouveure2019robot}. Contrary to the echo characteristic where only the target is directly scattered under open conditions, in such applications, the echo received by radar includes the signal reflected by terrain or buildings, referred to as multipath effects\cite{Lan2020joint,Kumar2020}. The presence of multipath signals will significantly impair radar's ability to detect and track targets (i.e., inducing false alarms and tracking target loss)\cite{Davey2019detection}. Thus, effectively managing and suppressing multipath effects induced by complex environments has become a challenging task\cite{LevyIsrael2023mcrb}. More recently, frequency-diverse array multiple input multiple output (FDA-MIMO) radar has garnered substantial attention\cite{Antonik2006frequency, Wang2015frequency, Xu2015joint}. Unlike phased array (PA) and MIMO radar systems, it introduces a small frequency increment between any two adjacent transmit elements, so that the transmit steering vector (SV) of FDA radar contains both target angle and distance information. Therefore, FDA-MIMO radar not only has the advantages of spatial diversity, but also has controllable degrees of freedom (DoF) in the distance dimension. Synthetically, its main applications include: parameter estimation\cite{Zhu2022cooperative,Lan2023adaptive}, target detection\cite{Huang2024robust},range clutter suppression\cite{Lan2021singlesnapshot}, range ambiguity resolution\cite{Wang2023resolving}, mainlobe interference suppression\cite{Lan2023beampattern,Lan2020suppression,Yang2023cognitive}, pattern synthesis\cite{Jia2023joint}, etc. Based on the frequency agility and waveform diversity characteristics between FDA-MIMO array elements, it can provide new ideas for multipath suppression. Therefore, this paper proposes a multipath identification and suppression method for FDA-MIMO radar to improve target detection capability and reduce false alarms.

In response to the multipath effect, researchers have conducted extensive explorations\cite{Yu2014mimo,Lan2023beampatterna,Tang2018multipath,Bialer2013maximumlikelihood,PallaresRodriguez2023dualpolarization,Wang2023superresolution,Luo2023effective,Liu2018height,Esposito2024deep,Xiang2020improved,Wu2023nlos,Yue2022cornerradar,Chen2023nonlineofsight,Gong2022multipathaided,Xiong2020time,Cheng2019temporal}. These studies can be broadly divided into two categories: $ \left. 1 \right)  $ is multipath mitigation or suppression. Such as adaptive antenna method\cite{Yu2014mimo,Lan2023beampatterna}, maximum likelihood method\cite{Tang2018multipath,Bialer2013maximumlikelihood}, polarization diversity technology\cite{PallaresRodriguez2023dualpolarization}, spatial diversity (e.g., MIMO) and high resolution technology\cite{Wang2023superresolution,Luo2023effective,Liu2018height}, neural network and machine learning\cite{Esposito2024deep,Xiang2020improved}. These methods aim to suppress multipath signals through algorithm or system design improvements, thus improving the accuracy of signal processing and the robustness of the system. $  \left. 2 \right)  $ is multipath utilization. Such as multipath perception of the building environment\cite{Wu2023nlos,Yue2022cornerradar}, non-line of sight target detection and tracking\cite{Chen2023nonlineofsight,Gong2022multipathaided}, time reversal\cite{Xiong2020time,Cheng2019temporal}, etc. Using the diffraction and multiple reflection signals formed by electromagnetic wave propagation, the multipath model is established and the signal processing method is improved to realize the detection of hidden targets, enhance the detection ability of weak targets and improve the capacity and accuracy of the communication system.

Due to the DoF brought by FDA-MIMO radar frequency increment and the waveform diversity characteristics, it can realize the joint estimation of distance and angle and provide the possibility of multipath suppression, which is different from traditional PA and MIMO. Currently, there are few studies on multipath interference suppression using FDA-MIMO radar. The main progress includes: In \cite{Zheng2022signal}, a generalized MUSIC algorithm is proposed to jointly estimate the angle and distance of low-elevation targets in meter-wave FDA-MIMO radar, but only the multipath model in single-target scenario is derived. In \cite{Liu2022discrimination,Liu2022robust}, a mainlobe deception target identification method for meter-wave FDA-MIMO radar is proposed, and the multipath effect of elevation angle in coherent source scene is considered. However, the computational complexity is high. In \cite{Cheng2019temporal}, the time focus of wireless communication transmission is achieved by combining FDA and time reversal. \cite{Cetintepe2014multipath} describes the multipath characteristics of FDA on the infinite earth plane.
However, these studies mainly focus on parameter estimation under multipath interference conditions, rarely analyze its advantages in multipath suppression from the characteristics of FDA, and rarely establish a general multipath analysis framework based on FDA-MIMO radar. 

This paper proposes a multipath signal identification and suppression method based on FDA-MIMO radar for target detection, which mainly solves the problem of specular reflection multipath of vehicle or UAV platform in urban or low-altitude environment of millimeter wave band, and mainly focuses on the situation that multipath component and direct path component are in different distance units and belong to specular reflection (i.e., the wavelength is much smaller than the roughness of the reflection surface and the distance resolution is high enough, which is common in millimeter wave broadband radar applications\cite{Cheng2019multipath}). We use FDA 's range-angle coupling characteristics to transform the multipath interference identification problem into a spatial spectrum estimation problem in the range cell to be detected. By employing the current range SV pertaining to the entity under observation to offset each snapshot data, one can separate the target radar echo associated with the current range unit from the range SV, whereas the echo of non-current range units remains coupled with the distance details. This strategy allows the achievement of accurate spatial spectrum data pertinent to the current range unit in the spatial spectrum estimation. That is, the covariance estimation is performed using the compensated data, and the corresponding spatial spectrum is estimated. If there are strong amplitude points with the same transmit and receive angles in the current spatial spectrum, it is considered that there is a real target in current test unit, otherwise it is a false target. In addition, in order to reduce the influence of second-order multipath (i.e., target 'ghost'\cite{Esposito2024deep}), we further reduce the echo intensity of multipath components by jointly optimizing the array transmit weighting and frequency increment\cite{ChengJoint2019}.
The main contributions in this paper include:
\begin{enumerate}
	\item Development of a comprehensive multipath signal receiving strategy for specular reflection conditions with FDA-MIMO radars for target detection. 
	\item Proposal for a multipath component identification method that is independent of prior environmental information and requires only monopulse echo. This technique also reduces multipath signal strength and subsequently reduces radar false alarms through transmission waveform and weight optimization.
\end{enumerate}

The rest of the paper is organized as follows: Section \ref{s2} introduces the system model including the signal and multipath model. Next, our method is proposed in Section \ref{s3}. The complexity analysis is in Section \ref{s4}. Numerical simulations of two scenarios are provided in Section \ref{s5}, and conclusions are finally drawn in Section \ref{s6}.

\textbf{\emph{Notations:}} $ \odot  $ represents the Hadamard product, $ \mathrm{diag}\left\{  \cdot  \right\}  $ denotes diagonalized  matrix. $ \left( \bullet \right) ^T $ and $ \left( \bullet \right) ^H $ denote the transpose and Hermitian transpose, respectively. $ \left\|  \cdot \right\| _2 $ represents the $ \ell _2 $ norm. $ E\left\{ \bullet \right\}  $ is the statistical expectation value. $ \left( \bullet \right) ^* $ implies complex conjugation. $ \mathbf{I} $ denotes the identity matrix of the corresponding order. $ \mathbf{1}_{M} $ denotes $ M $ dimensional unity vector. $ \mathbb{Z},\mathbb{C} $ stands for the integral and complex numbers, respectively.		
\section{SYSTEM MODEL}
\label{s2}
\subsection{signal model}
\begin{figure}[htbp]
	\centering
	\subfigure {\includegraphics[width=0.5\textwidth]{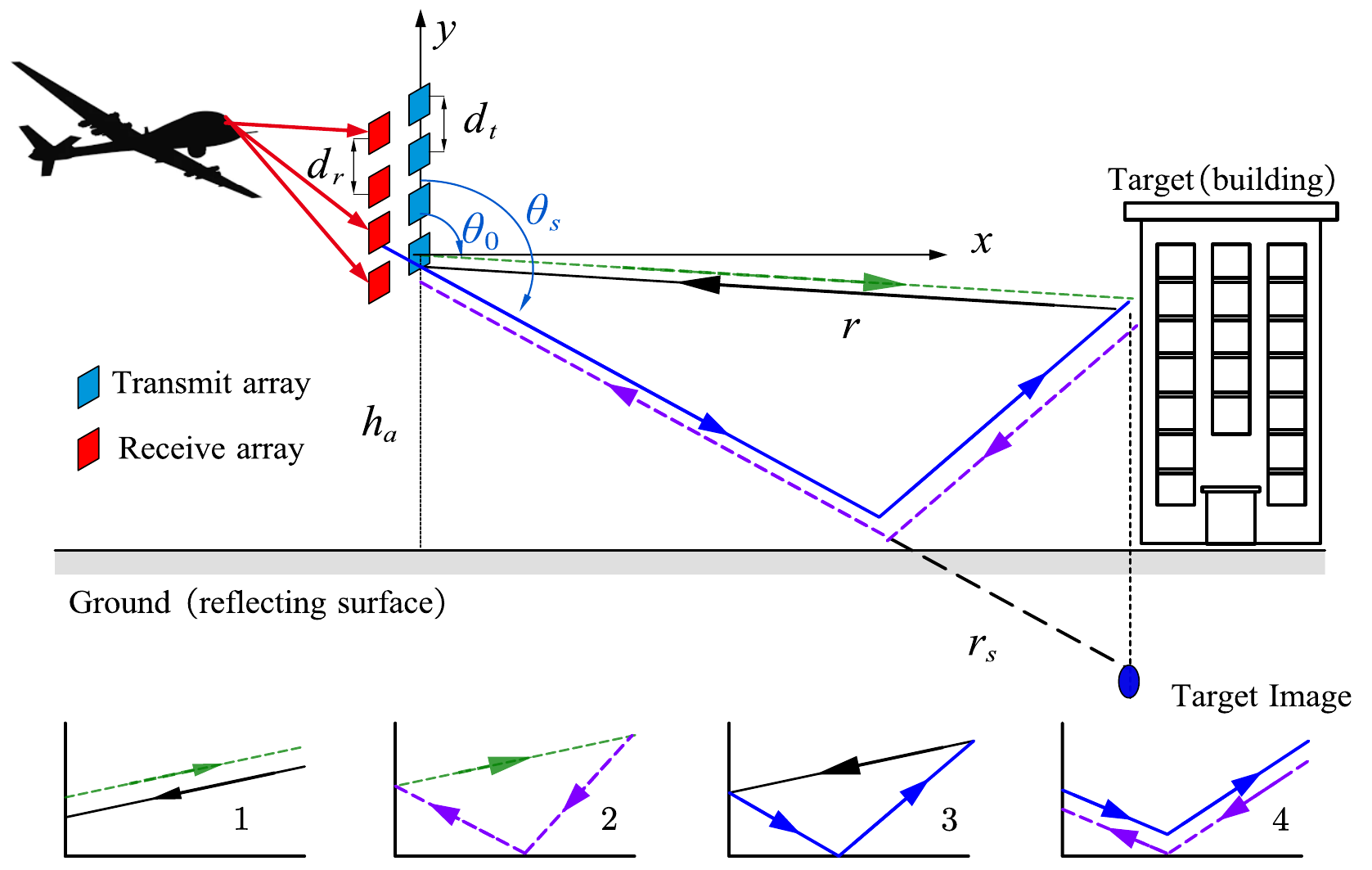}}
	\caption{Model of FDA-MIMO radar which is installed on a UAV in urban areas, including four transmission paths of electromagnetic wave propagation, which are 1) directed –directed wave, 2) directed – reflected wave, 3) reflected–directed wave, 4) reflected– reflected wave.}
	\label{f.1}
\end{figure}

Considering a monostatic FDA-MIMO radar installed on a UAV for target detection application in urban multipath environment, the array consists of $ M $ transmit elements and $ N $ receive elements, which is a linear array, as shown in Fig. \ref{f.1}. Then, suppose a far-field point target with angle-range pair $ \left( r , \theta  \right) $, where $ \theta $ is the azimuth angle, and $ r $ represents the range between the target and the origin point. The carrier frequencies of each element are $ f_m=f_0+\left( m-1 \right) \Delta f ,m=1,\cdots,M$, and $ \Delta f  $ is the frequency increment of FDA-MIMO radar. According to the principle of geometric optics, there are four transmission paths for a far-field point target, as shown in Fig. \ref{f.1}. For the non-direct transmission path 4 ( the lower right corner of Fig. \ref{f.1}), the reflected path range can be represented as
\begin{equation}\label{1}
	r_s=\sqrt{\left( r\sin \left( \theta \right) \right) ^2+\left( 2h_a+r\cos \left( \theta \right) \right) ^2}
\end{equation}
where $ h_a $ represents the projection distance between the array and the reflector (ground). The azimuth angle of the reflected path can be represented as
\begin{equation}\label{2}
	\theta _s=\pi -\tan ^{-1}\left( \frac{r\sin \left( \theta \right)}{2h_a+rcos \left( \theta \right)} \right) 
\end{equation}
Therefore, these reflection path echoes can be regarded as the echo of a mirror target $ \left( r_s,\theta _s \right)  $ on the surface of the reflector.
Assume the transmitted radio signals are $ \mathbf{x}\left( t \right) =\left[ x_1\left( t \right) ,\cdots ,x_M\left( t \right) \right] ^T $, and the corresponding baseband signal are $ \mathbf{s}\left( t \right) =\left[ s_1\left( t \right) ,\cdots ,s_M\left( t \right) \right] ^T $, $ x_m\left( t \right) =s_m\left( t \right) e^{j2\pi f_mt} ,m =1,\cdots,M$, which are orthogonal to each other that
\begin{equation}\label{key}
	\int_0^{T_s}{}s_m\left( t \right) s_{m'}^{*}\left( t \right) dt=\left\{ \begin{array}{c}
		1,m=m'\\
		0,m\ne m'\\
	\end{array} \right. 
\end{equation}
where $T_s  $ is the pulse duration. It is assumed that the electromagnetic wave propagates independently in free space. Thus, for the direct path target $ \left( \theta  , r \right) $, the received echo of the $ n $-th element ($ n=1,\cdots,N $) can be written as
\begin{equation}\label{4}
	\begin{split}
		y_n\left( t \right)& =\sum_{m=1}^M{}\sqrt{\frac{P_t}{M}}\tilde{\eta}_0s_m\left( t-\tau _0 \right) e^{j2\pi \left( f_m+f_{d,m} \right) \left( t-\tau _0 \right)}\cdot 
		\\
		&\qquad \qquad e^{j2\pi \frac{d_t}{\lambda _0}\left( m-1 \right) \cos \theta}e^{j2\pi \frac{d_r}{\lambda _0}\left( n-1 \right) \cos \theta}
	\end{split}
\end{equation}
where $ d_t $ and $ d_r $ are the interspacings of transmit and receive antennas, respectively. $ \lambda_0=c/f_0 $ is the reference wavelength and $ c $ is the optical velocity. $ \tilde{\eta}_0 $ is the complex scattering coefficient of the target. $ \tau_0=2r/c $. $ P_t $ is the total transmit power. Here we assume that the transmit power is evenly distributed among the transmit elements. $ f_{d,m}$ represents the Doppler frequency with respect to $ m $-th transmit waveform, i.e., $ f_{d,m}=2v_t/\lambda_m $, $ v_t $ is the radial velocity of the target, $ \lambda_m=c/f_m $ is the wavelength corresponding to the $ m $-th transmit element.
\begin{figure}[htbp]
	\centering
	\subfigure {\includegraphics[width=0.45\textwidth]{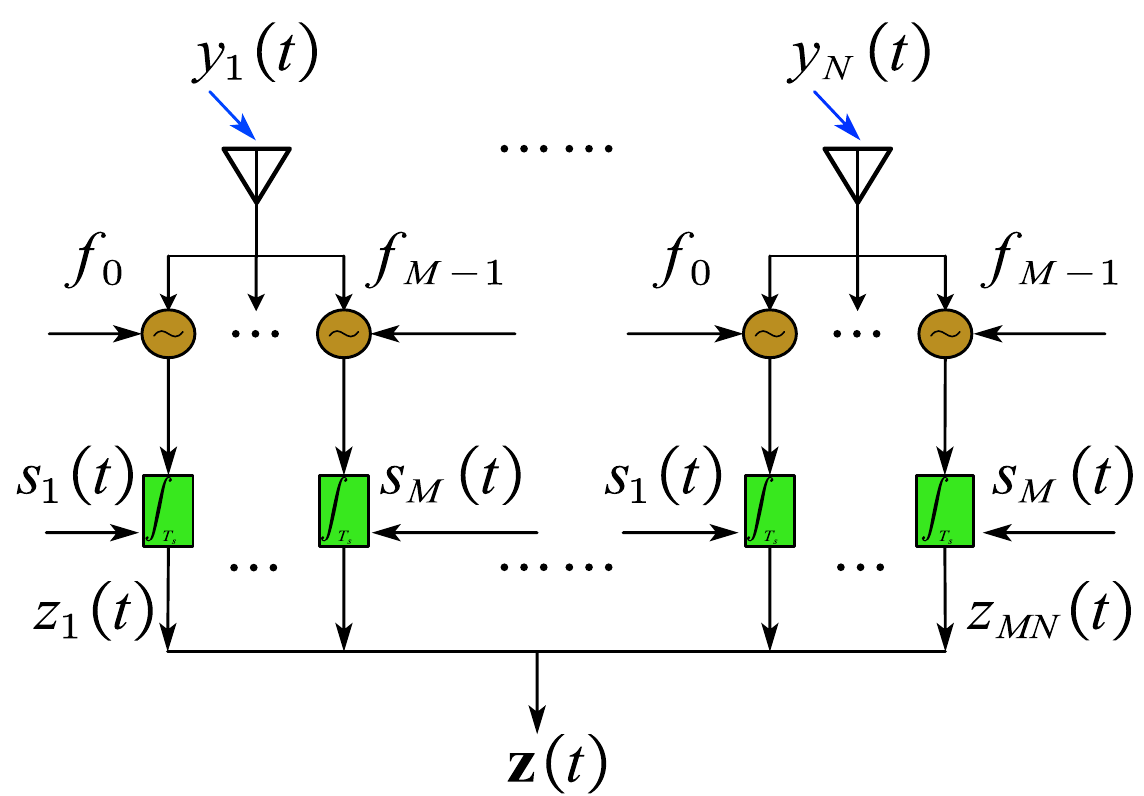}}
	\caption{FDA-MIMO radar receiver based on multi-channel mixing and matched filtering.}
	\label{f.2}
\end{figure}
Fig. \ref{f.2} illustrates the general FDA-MIMO receiver structure proposed in \cite{Gui2018coherent}. The echo signal is initially mixed using multiple mixers, resulting in the generation of a multi-channel signal. After matching filtering, the $ m $-th output of the $ n $-th received element is
\begin{equation}\label{5}
	z_{n,m}\left( t \right) =\int{}y_n\left( t' \right) e^{-j2\pi f_mt'}s_{m}^{*}\left( t-t' \right) dt'
\end{equation}
where $ t' $ is the integration variable. Then, substituting \eqref{4} into \eqref{5}, we obtain
\begin{equation}\label{key}
	\begin{split}
		&z_{n,m}\left( t \right) =\sqrt{\frac{P_t}{M}}\tilde{\eta}_0a_{R,n}\sum_{m'=1}^M{}a_{T,m'}e^{-j2\pi f_{m'}\tau _0}e^{-j2\pi f_{d,m'}\tau _0}\cdot 
		\\
		&\int{}\left[ s_{m'}\left( t'-\tau _0 \right) \cdot s_{m}^{*}\left( t-t' \right) e^{j2\pi \left( f_{m'}+f_{d,m'}-f_m \right) t'} \right] dt'
		\\
		&=\sqrt{\frac{P_t}{M}}\tilde{\eta}_0a_{R,n}e^{-j2\pi f_1\tau _0}\sum_{m'=1}^M{}a_{T,m'}e^{-j2\pi \left( m'-1 \right) \Delta f\tau _0}\cdot 
		\\
		&e^{-j2\pi f_{d,m'}\tau _0}\chi _{m,m'}\left( t-\tau _0,\left[ m'-m \right] \Delta f+f_{d,m'} \right) 
	\end{split}
\end{equation}
For notation simplicity, we have
\begin{equation}\label{key}
	\begin{aligned}
		a_{R,n}&=e^{j2\pi \frac{d_r}{\lambda _0}\left( n-1 \right) \cos \theta},n=1,\cdots ,N
		\\
		a_{T,m}&=e^{j2\pi \frac{d_t}{\lambda _0}\left( m-1 \right) \cos \theta},m=1,\cdots ,M
	\end{aligned}
\end{equation}
And $ \chi _{m',m} $ is the ambiguity function of baseband waveform $ s_m(t) $ and $ s_{m'}\left( t \right)  $, which defines as
\begin{equation}\label{key}
	\chi _{m',m}\left( \Delta t,f \right) \triangleq \int_{-\infty}^{\infty}{}s_{m'}\left( t \right) s_{m}^{*}\left( t-\Delta t \right) e^{-j2\pi ft}dt
\end{equation}
Then, the filtering signal samples of $ MN $ channels are collected in the data matrix $ \mathbf{Z}\left( t \right) \in \mathbb{C} ^{N\times M} $, that is
\begin{equation}\label{9}
	\mathbf{Z}\left( t \right) =\eta _0\mathbf{a}_r\left( \theta \right) \left( \mathbf{a}_t\left( \theta \right) \odot \mathbf{\Gamma }\left( r \right) \odot \mathbf{e}_{fd}\left( \tau _0 \right) \right) ^T\mathbf{R}_{ss}\left( t-\tau _0 \right) 
\end{equation}
where $ \eta _0=\sqrt{\frac{P_t}{M}}\tilde{\eta}_0e^{-j2\pi f_1\tau _0} $ is the equivalent coefficient. $ \mathbf{a}_r\left( \theta \right) \in \mathbb{C} ^{N\times 1} $ and $ \mathbf{a}_t\left( \theta \right) \in \mathbb{C} ^{M\times 1} $ are the receive and transmit angular SV respectively, and $ \mathbf{\Gamma }\left( r \right) $ is the transmit range SV, which can be denoted as 
\begin{subequations}\label{key}
	\begin{align}
		\mathbf{a}_r\left( \theta \right) &=\left[ 1,e^{j2\pi \frac{d_r}{\lambda _0}\cos \theta},\cdots ,e^{j2\pi \frac{d_r}{\lambda _0}\left( N-1 \right) \cos \theta} \right] ^T \\
		\mathbf{a}_t\left( \theta \right) &=\left[ 1,e^{j2\pi \frac{d_t}{\lambda _0}\cos \theta},\cdots ,e^{j2\pi \frac{d_t}{\lambda _0}\left( M-1 \right) \cos \varphi} \right] ^T\\
		\mathbf{\Gamma }\left( r \right)& =\left[ 1,e^{-j2\pi \frac{2\Delta f}{c}r},\cdots ,e^{-j2\pi \frac{2\Delta f}{c}\left( M-1 \right) r} \right] ^T
	\end{align}
\end{subequations}
and $ \mathbf{e}_{fd}\left( \tau _0 \right) $ is the Doppler delay which can be denoted by
\begin{equation}\label{key}
	\mathbf{e}_{fd}\left( \tau _0 \right) =\left[ e^{-j2\pi f_{d,1}\tau _0},\cdots ,e^{-j2\pi f_{d,M}\tau _0} \right] ^T
\end{equation}
$ \mathbf{R}_{ss}\left( t-\tau _0 \right) \in \mathbb{C} ^{M\times M} $ is the ambiguity matrix of FDA-MIMO radar transmit waveform with frequency increment and distinct delay which can be denoted by
\begin{equation}\label{key}
	\left[ \mathbf{R}_{ss}\left( t-\tau _0 \right) \right] _{m,m'}=\chi _{m,m'}\left( t-\tau _0,\left[ m'-m \right] \Delta f+f_{d,m'} \right) 
\end{equation}
Considering the echo signal structure from the perspective of MIMO radar virtual array, \eqref{9} can be rewritten as a vector $\mathbf{z}\left( t \right) \in \mathbb{C} ^{NM\times 1}  $, i.e.,
\begin{equation}\label{key}
	\begin{split}
		\mathbf{z}\left( t \right) &=\eta _0\left[ \mathbf{I}_N\otimes \left( \mathrm{diag}\left( \mathbf{e}_{fd}\left( \tau _0 \right) \right) \mathbf{R}_{ss}^{T}\left( t-\tau _0 \right) \right) \right] \cdot 
		\\
		&\qquad\qquad \left[ \mathbf{a}_r\left( \theta \right) \otimes \mathbf{a}_t\left( r,\theta \right) \right] 
	\end{split}
\end{equation}
where $ \mathbf{a}_t\left( r,\theta \right) =\mathbf{a}_t\left( \theta \right) \odot \mathbf{\Gamma }\left( r \right)  $ is the range-angle-dependent transmit SV\cite{Huang2024robust}.
\subsection{multipath model}
Then considering the multipath environment, the echo vector $ \mathbf{y}\left( t \right) =\left[ y_1\left( t \right) ,\cdots ,y_N\left( t \right) \right] \in \mathbb{C} ^{N\times 1} $ is the adding of the four paths shown in Fig.\ref{f.1} with all the $ M $ transmitted signals:
\begin{equation}\label{13}
	\mathbf{y}\left( t \right) =\eta _0\left[ \begin{array}{c}
		\mathbf{a}_r\left( \theta \right) \mathbf{a}_{t}^{T}\left( \theta \right) \mathbf{x}\left( t-2r/c \right)\\
		+\tilde{\rho}_0\mathbf{a}_r\left( \theta _s \right) \mathbf{a}_{t}^{T}\left( \theta \right) \mathbf{x}\left( t-\left( r+r_s \right) /c \right)\\
		+\tilde{\rho}_0\mathbf{a}_r\left( \theta \right) \mathbf{a}_{t}^{T}\left( \theta _s \right) \mathbf{x}\left( t-\left( r+r_s \right) /c \right)\\
		+\tilde{\rho}_{0}^{2}\mathbf{a}_r\left( \theta _s \right) \mathbf{a}_{t}^{T}\left( \theta _s \right) \mathbf{x}\left( t-2r_s/c \right)\\
	\end{array} \right] +\mathbf{v}\left( t \right) 
\end{equation}
where $ \tilde{\rho}_0 $ are the complex scattering coefficient of the reflector, $ \mathbf{v}\left( t \right) \in \mathbb{C} ^{N\times 1} $ is the independent noise which is assumed to be a zero-mean white circularly Gaussian vector with covariance $ \sigma _{n}^{2}\mathbf{I}_N $. The Doppler term $ e^{j2\pi f_{d,m}\left( t \right)} $ in \eqref{4} is absorbed into the transmit signal $\mathbf{x}\left( t \right)   $  in \eqref{13} for the sake of simplicity.

For further derivations of the multiple channel signal from \eqref{13}, we introduce the following assumptions:
\begin{enumerate}
	\item {\textbf{Far-field Hypothesis}}: The target range is far enough that $ r\gg 2D^2/\lambda_0 $, where $ D $ is the array aperture. 
	\item {\textbf{Bandwidth Hypothesis}}: The bandwidth of each baseband waveform $ s_m(t) $ denoted by $ B_m $ satisfies the condition that $ 2vT_s/c\ll 1/B_m$. Thus, we obtain $ \Delta f\ll f_0 $, so that the Doppler frequency shift of $ M $ receiving channels is approximately equal, i.e., $ \mathbf{e}_{fd}\left( \tau _0 \right) \approx e^{-j4\pi v_t\tau _0f_0/c}\mathbf{1}_{M}^{T} $. 
\end{enumerate}
Using the aforementioned receiver and these assumptions, the $ MN $ channel signal is
\begin{equation}\label{15}
	\begin{split}
	 	\mathbf{z}\left( t \right)&=\eta _0\left\{ \left[ \mathbf{I}_N\otimes \mathbf{R}_{ss}^{T}\left( t-2r/c \right) \right] \cdot \left[ \mathbf{a}_r\left( \theta \right) \otimes \mathbf{a}_t\left( r,\theta \right) \right] \right. 
		\\
		&+\tilde{\rho}_0\left[ \mathbf{I}_N\otimes \mathbf{R}_{ss}^{T}\left( t-2\bar{r}/c \right) \right] \cdot \left[ \mathbf{a}_r\left( \theta _s \right) \otimes \mathbf{a}_t\left( \bar{r},\theta \right) \right] 
		\\
		&+\tilde{\rho}_0\left[ \mathbf{I}_N\otimes \mathbf{R}_{ss}^{T}\left( t-2\bar{r}/c \right) \right] \cdot \left[ \mathbf{a}_r\left( \theta \right) \otimes \mathbf{a}_t\left( \bar{r},\theta _s \right) \right] 
		\\
		&\left. +\tilde{\rho}_{0}^{2}\left[ \mathbf{I}_N\otimes \mathbf{R}_{ss}^{T}\left( t-2r_s/c \right) \right] \cdot \left[ \mathbf{a}_r\left( \theta _s \right) \otimes \mathbf{a}_t\left( r_s,\theta _s \right) \right] \right\} \\
		&+\breve{\mathbf{v}}\left( t \right) 
	\end{split}
\end{equation}
where $ \mathbf{z}\left( t \right) \in \mathbb{C} ^{MN\times 1}  $ and the Doppler frequency shift $ e^{-j4\pi v_t\tau _0f_0/c} $ absorbs into the reflection coefficient $ \eta _0 $. $ \bar{r}=\left( r+r_s \right) /2 $ is the equivalent distance of the first-order multipath component. $ \breve{\mathbf{v}}\left( t \right)  $ is the noise vector after multi-channel matched filtering, and the covariance of it is
\begin{equation}\label{key}
\begin{split}
			\mathbf{R}_{\breve{\mathbf{v}}}&=\left[ \mathbf{1}_N\otimes \mathbf{s}\left( t \right) \right] \left[ \mathbf{1}_N\otimes \mathbf{s}\left( t \right) \right] ^H\\
		&=\left( \mathbf{1}_N\mathbf{1}_{N}^{T} \right) \otimes \mathbf{R}_{ss}\left( 0 \right) 
\end{split}
\end{equation}

As shown in \eqref{15}, due to the presence of multipath components in the echo signal, a peak will appear not only at the real target range $ r $, but also at the multipath component range $ r_s $ after the matched filtering. As a result, false alarms occur, affecting target detection and tracking capabilities. Therefore, in the following we use the range-angle coupling characteristics of FDA-MIMO to distinguish the real target after matched filtering, so as to suppress the multipath component.

\section{PROPOSED METHOD}
\label{s3}
\subsection{Multipath Discrimination}
It can be seen from Fig.\ref{f.1} that the direct path and the non-direct path are different in the transmitting space angle and the receiving space angle. We can distinguish the real target and the false target generated by the first-order multipath echo from the transmitting angle domain and the receiving angle domain. Therefore, we propose the following multipath component identification method.
First, we define the transmit spatial frequency $ f_{st} $ and receive spatial frequency $ f_{sr} $ as \cite{Xu2015joint}, 
\begin{subequations}\label{}
	\begin{align}
		f_{st}&=-\frac{2\Delta f}{c}r+\frac{d_t}{\lambda _0}\cos \theta 
		\\
		f_{sr}&=\frac{d_r}{\lambda _0}\cos \theta 
	\end{align}
\end{subequations}
For FDA-MIMO radar, after range dependence compensation, the transmit spatial frequency is a linear function of the received spatial frequency\cite{Xu2015joint}. Particularly, if the interspaces of the transmit and receive antennas are the same, i.e., $ d_t=d_r $, the real target spectrum is diagonally distributed in the joint transmit-receive spatial frequency domains while the multipath component is not. 
Thus, using the distance information of the current unit to be estimated, the compensating vector in the joint transmit-receive spatial frequency domains can be defined as
\begin{equation}\label{18}
	\mathbf{g}(r)=\mathbf{1}_N\otimes \mathbf{\Gamma }^*\left( r \right) 
\end{equation}
where $ \mathbf{\Gamma }^*\left( r \right)  $ represents the conjugate of the transmit range SV. Then, after compensation, the received data can be expressed as 
\begin{equation}\label{19}
	\begin{split}
		&	\mathbf{z}_{cp}\left( t;r \right) =\mathbf{z}\left( t \right) \odot \mathbf{g}\left( r \right) =\mathrm{diag}\left( \mathbf{g} \left( r \right) \right) \mathbf{z}\left( t \right) 
		\\
		&=\eta _0\left\{ \left[ \mathbf{I}_N\otimes \mathbf{R}_{ss}^{T}\left( t-2r/c \right) \right] \cdot \left[ \mathbf{a}_r\left( \theta \right) \otimes \mathbf{a}_t\left( \theta \right) \right] \right. 
		\\
		&+\tilde{\rho}_0\left[ \mathbf{I}_N\otimes \mathbf{R}_{ss}^{T}\left( t-2\bar{r}/c \right) \right] \cdot \left[ \mathbf{a}_r\left( \theta _s \right) \otimes \mathbf{a}_t\left( \bar{r}-r,\theta \right) \right] 
		\\
		&	+\tilde{\rho}_0\left[ \mathbf{I}_N\otimes \mathbf{R}_{ss}^{T}\left( t-2\bar{r}/c \right) \right] \cdot \left[ \mathbf{a}_r\left( \theta \right) \otimes \mathbf{a}_t\left( \bar{r}-r,\theta _s \right) \right] 
		\\
		&	\left. +\tilde{\rho}_{0}^{2}\left[ \mathbf{I}_N\otimes \mathbf{R}_{ss}^{T}\left( t-2r_s/c \right) \right] \cdot \left[ \mathbf{a}_r\left( \theta _s \right) \otimes \mathbf{a}_t\left( r_s-r,\theta _s \right) \right] \right\} 
		\\
		&	+\breve{\mathbf{v}}\left( t \right) \odot \mathbf{g}\left( r \right) 
	\end{split}
\end{equation}
Note that it has no effect on the spectral distribution of the compensated noise $ \breve{\mathbf{v}}\left( t \right) \odot \mathbf{g} $\cite{Xu2015joint}. Thus, the covariance remains unchanged for $ \breve{\mathbf{v}}\left( t \right)  $.
Due to the existence of component $ \left[ \mathbf{I}_N\otimes \mathbf{R}_{ss}^{T}\left( t-2r/c \right) \right] \cdot \left[ \mathbf{a}_r\left( \theta \right) \otimes \mathbf{a}_t\left( r,\theta \right) \right]  $, the real single-path target will appear on the diagonal of the joint transmit-receive spatial frequency domain, while the remaining multipath signals will appear on the non-diagonal line due to the wrong distance compensation. 
For comparison, we write the received data for MIMO radar here. Note that we only need to set $\Delta f=0  $ in \eqref{19}, i.e.,
\begin{equation}\label{key}
	\begin{split}
	&	\mathbf{z}_{MIMO}\left( t \right) =\eta _0\left\{ \left[ \mathbf{I}_N\otimes \mathbf{R}_{ss}^{T}\left( t-2r/c \right) \right] \cdot \left[ \mathbf{a}_r\left( \theta \right) \otimes \mathbf{a}_t\left( \theta \right) \right] \right. 
		\\
	&\qquad	\qquad\quad+\tilde{\rho}_0\left[ \mathbf{I}_N\otimes \mathbf{R}_{ss}^{T}\left( t-2\bar{r}/c \right) \right] \cdot \left[ \mathbf{a}_r\left( \theta _s \right) \otimes \mathbf{a}_t\left( \theta \right) \right] 
		\\
	&\qquad	\qquad\quad+\tilde{\rho}_0\left[ \mathbf{I}_N\otimes \mathbf{R}_{ss}^{T}\left( t-2\bar{r}/c \right) \right] \cdot \left[ \mathbf{a}_r\left( \theta \right) \otimes \mathbf{a}_t\left( \theta _s \right) \right] 
		\\
	&	\left. +\tilde{\rho}_{0}^{2}\left[ \mathbf{I}_N\otimes \mathbf{R}_{ss}^{T}\left( t-2r_s/c \right) \right] \cdot \left[ \mathbf{a}_r\left( \theta _s \right) \otimes \mathbf{a}_t\left( \theta _s \right) \right] \right\} +\breve{\mathbf{v}}\left( t \right) 
	\end{split}
\end{equation}
It is worth noting that we only obtain separated spatial frequency domain and range domain information for MIMO radar (through matched filtering), while joint spatial frequency and range information is obtained using FDA-MIMO. In other words, MIMO radar gives information either from the range or angle domain that we cannot distinguish the real target from the multipath component after matching filtering. Moreover, since the aspects proposed in this paper mainly identify multipath targets from the spatial domain, the motion state of the target has little effect on the identification results.

According to the Capon spectrum theorem \cite{Jakobsson2000computationally}, assume that the range results of constant false alarm rate (CFAR) detection \cite{Guerci2015spacetime} after matched filtering is $ \mathcal{R} =\left\{ r_1,\cdots ,r_L \right\}  $, where $r_i,i=1,\cdots ,L$ is the $ i $-th suspected target unit to be detected. Then, the estimated angles of the transmit and receive spatial domain can be obtained by
\begin{equation}\label{21}
	\left( \hat{\theta}_t,\hat{\theta}_r \right) =\underset{\theta _t,\theta _r}{\mathrm{arg}\max}\left\| \mathbf{w}_{}^{H}\left( \theta _t,\theta _r \right) \mathbf{z}_{cp}\left( r_i/c \right) \right\| _1
\end{equation}
where $ 		\mathbf{w}\left( \theta _t,\theta _r \right) =\mathbf{Q}_{cp}^{-1}\left[ \mathbf{a}_r\left( \theta _r \right) \otimes \mathbf{a}_t\left( \theta _t \right) \right]  $ and $ 	\mathbf{Q}_{cp} $ is the covariance matrix after compensated by the current distance to be detected, i.e.,
\begin{equation}\label{22}
	\mathbf{Q}_{cp}^{}=\frac{1}{T_s}\int_0^{T_s}{}\mathbf{z}_{cp}\left( t;r_i \right) \mathbf{z}_{cp}^{H}\left( t;r_i \right) dt
\end{equation}
If $ \hat{\theta}_t=\hat{\theta}_r $, we can infer that the current range cell to be detected is the real target under the intrinsic false alarm rate, otherwise it is not.
To summarize, the algorithm of multipath discrimination is shown in Algorithm \ref{alg1}.
\begin{algorithm}[htb] 
	\caption{Proposed multipath discrimination algorithm for FDA-MIMO} 
	\label{alg1} 
	\begin{algorithmic}[1] 
		\Require
		\\
		Receive signal after matched filtering: $\mathbf{z}\left( t \right)  $; 
		\\
		CFAR detection results: $ \mathcal{R} =\left\{ r_1,\cdots ,r_L \right\}  $;
		\Ensure
		
		Multipath identification results: $ \bar{\mathcal{R}}=\left\{ 1_{\left( 1 \right)},\cdots ,0_{\left( L \right)} \right\} $, where $ 1 $ indicates that the current range cell has a real target, and $ 0 $ indicates the opposite;
		\item Initial index $ i=1 $;
		\item Obtaining the range compensation vector:$ \mathbf{g}\left( r_i \right)  $, and using (\ref{19}) to obtain the compensated signal vector: $\mathbf{z}_{cp}\left( t;r \right) $; 	\label{code11}
		\item Using \eqref{21} to obtain transmit-receive spatial domain estimation set $ \left( \hat{\theta}_t,\hat{\theta}_r \right)  $; 
		\label{code222}
		\item if $ \hat{\theta}_t=\hat{\theta}_r $, 
			
	$ 1_{\left( i \right)} $ is the CFAR detection result.
		
else, 
		 
		 $ 0_{\left( i \right)} $ is the CFAR detection result.
		\label{code32}
		\item Repeat step \ref{code11}, \ref{code222} and \ref{code32} until $ i>L $.\\
		
		\Return :$ \bar{\mathcal{R}} $. 
	\end{algorithmic}
\end{algorithm} 

\subsection{Multipath Mitigation}
After obtaining the range-angle information of the real target, we can design a multipath mitigation scheme based on the auto-scanning characteristics of FDA-MIMO radar\cite{Wang2022frequency}.
Our focus here is to maximize the energy delivered to the target space while minimizing the energy radiated in the multipath direction. To achieve this goal, we can optimize the transmit waveform parameters (frequency increment), as well as transmit and receive weights. Thus, the optimization problem can be expressed as
\begin{equation}\label{key}
	\underset{\mathbf{w}_R,\mathbf{w}_F,\Delta f}{\max}f\left( \mathbf{w}_{R}^{},\mathbf{w}_{F}^{},\Delta f \right) =\frac{\left| \mathbf{w}_{R}^{H}\mathbf{z}\left( \tau _0 \right) \right|^2}{\mathbf{w}_{R}^{H}\mathbf{R}_{zz}^{}\mathbf{w}_{R}^{}}
\end{equation}
where $ f\left( \mathbf{w}_{R}^{},\mathbf{w}_{F}^{},\Delta f \right)  $ represents the signal-to-interference-plus-noise ratio (SINR) at the receiver. $ \mathbf{w}_{R}^{}\in \mathbb{C} ^{MN\times 1} $ and $ \mathbf{w}_{F}^{}\in \mathbb{C} ^{M\times 1} $ are the receive and transmit weighting, respectively. $ \mathbf{R}_{zz}$ is the received covariance matrix which is usually calculated by discrete sum, i.e.,
\begin{equation}\label{24}
	\mathbf{R}_{zz}^{}=\frac{1}{T_s}\int_0^{T_s}{\mathbf{z}\left( t;\Delta f,\mathbf{w}_{F}^{} \right) \mathbf{z}^H\left( t;\Delta f,\mathbf{w}_{F}^{} \right)}dt
\end{equation}
It can be transformed to a constraint minimize problem, i.e.,
\begin{equation}\label{25}
	\begin{cases}
		\underset{\mathbf{w}_R,\mathbf{w}_F,\Delta f}{\min}\mathbf{w}_{R}^{H}\mathbf{R}_{zz}^{}\mathbf{w}_{R}^{}\\
		\,\quad	s.t. \,\,\quad\mathbf{w}_{R}^{H}\breve{\mathbf{a}}_{r,t}\left( \theta ;r,\theta \right) =1\\
	\end{cases}
\end{equation}
where $\breve{\mathbf{a}}_{r,t}\left( \theta ;r,\theta \right) =\mathbf{a}_r\left( \theta \right) \otimes \breve{\mathbf{a}}_t\left( r,\theta \right) $ is visual  transmit-receive SV and $ \breve{\mathbf{a}}_t\left( r,\theta \right) =\mathbf{w}_F\odot \mathbf{a}_t\left( r,\theta \right)  $ is the equivalent transmit SV which considering the transmit weighting. Note that \eqref{25} is a multi-variable optimization problem. 
\subsubsection{Receive weighting}
In practice, the receiving and transmitting weights and waveform parameters are designed separately. Thus, we can assume that the optimal receive weighting is obtained when the transmit weighting and frequency increment are known. Obviously, \eqref{25} is a minimum power distortionless response (MPDR) problem \cite{Li2006} in this case, and the solution is well known as 
\begin{equation}\label{26}
	\mathbf{w}_{R}=\xi \mathbf{R}_{zz}^{-1}\breve{\mathbf{a}}_{r,t}\left( \theta ;r,\theta \right) 
\end{equation}
where $\xi =\left\{ \breve{\mathbf{a}}_{r,t}^{H}\left( \theta ;r,\theta \right) \mathbf{R}_{zz}^{-1}\breve{\mathbf{a}}_{r,t}\left( \theta ;r,\theta \right) \right\} ^{-1} $ is the coefficient.
\subsubsection{Transmit weighting}
Substituting \eqref{26} into \eqref{25}, we obtain an unconstrained minimization problem
\begin{equation}\label{27}
	\underset{\mathbf{w}_F,\Delta f}{\min}\left| \xi \right|^2\breve{\mathbf{a}}_{r,t}^{H}\left( \theta ;r,\theta \right) \mathbf{R}_{zz}^{-H}\breve{\mathbf{a}}_{r,t}\left( \theta ;r,\theta \right) 
\end{equation}
Using the expressing of $ \xi $, \eqref{27} can be transformed into
\begin{equation}\label{28}
	\begin{cases}
		\underset{\mathbf{w}_F,\Delta f}{\max}\,\,\breve{\mathbf{a}}_{r,t}^{H}\left( \theta ;r,\theta \right) \mathbf{R}_{zz}^{-1}\breve{\mathbf{a}}_{r,t}\left( \theta ;r,\theta \right)\\
		s.t.   \qquad \left\| \mathbf{w}_F \right\| _2=\sqrt{M}\\
	\end{cases}
\end{equation}
For the same reason that we solve $ \mathbf{w}_{R}$, we assume that the optimal frequency increment $\Delta_f  $ is known. Then the closed-form solution of \eqref{28} can be obtained (the detailed derivation process is shown in Appendix \ref{a1}), i.e.,
\begin{equation}\label{29}
	\mathbf{w}_F=\frac{\sqrt{M}}{\left\| \left( \boldsymbol{\nu }_1 \right) _M \right\| _{2}^{2}}\mathbf{a}_t\left( r,\theta \right) \left( \boldsymbol{\nu }_1 \right) _M\left( \boldsymbol{\nu }_1 \right) _{M}^{H}
\end{equation} 
where $ \left( \boldsymbol{\nu }_1 \right) _M \in \mathbb{C} ^{M\times 1}$ is the eigenvector corresponding to the maximum eigenvalue of $ \mathbf{R}_{zz}^{} $ which takes the first $ M $ values, i.e.,
\begin{equation}\label{30_0}
	\begin{split}
		\mathbf{R}_{zz}^{}&=\mathbf{V\Sigma }_R\mathbf{V}^H
		\\
		\mathbf{\Sigma }_R&=\mathrm{diag}\left( \varrho _1,\cdots ,\varrho _{MN} \right) 
	\end{split}
\end{equation}
where $ \mathbf{VV}^H=\mathbf{I},\mathbf{V}=\left[ \boldsymbol{\nu }_1,\cdots ,\boldsymbol{\nu }_{NM} \right] \in \mathbb{C} ^{NM\times NM}, $ is the characteristic matrix of $  \mathbf{R}_{zz}$, and $ \varrho _1\geqslant \varrho _2\geqslant \cdots \geqslant \varrho _{MN} $ are the eigenvalues.
\subsubsection{frequency increment}
After obtaining the transmit weighting, we focus on waveform optimization design. In general, for MIMO radar, there are many kinds of orthogonal waveforms to choose from, and interested readers can refer to \cite{Majumder2016design}. However, in this paper, we mainly discuss FDA-MIMO radar. Therefore, for simplicity, we only have the waveform design freedom of frequency increment, and the optimal frequency offset satisfying the multipath suppression condition is obtained by the following optimization problem:
\begin{equation}\label{31}
	\underset{\Delta f}{\max}\,\,\psi \left( \Delta f \right) =\breve{\mathbf{a}}_{r,t}^{H}\left( \theta ;r,\theta \right) \mathbf{R}_{zz}^{-1}\breve{\mathbf{a}}_{r,t}\left( \theta ;r,\theta \right) 
\end{equation}
Then, it can be deduced as
\begin{subequations}\label{30}
	\begin{align}
		\psi \left( \Delta f \right) &=\mathrm{tr}\left( \mathbf{R}_{zz}^{-1}\breve{\mathbf{a}}_{r,t}\left( \theta ;r,\theta \right) \breve{\mathbf{a}}_{r,t}^{H}\left( \theta ;r,\theta \right) \right) \label{30_a}
		\\
		&=\mathrm{tr}\left( \mathbf{R}_{zz}^{-1}\mathbf{U\Sigma }_a\mathbf{U}^H \right) \label{30_b}
		\\
		&=\mathrm{tr}\left( \mathbf{U}^H\mathbf{R}_{zz}^{-1}\mathbf{U\Sigma }_a \right) =\mathrm{tr}\left( \mathbf{\Psi \Sigma }_a \right) \label{30_c}
		\\
		&=\sum_{i=1}^{MN}{\Psi _{ii}\sigma _i}=\sigma _1\Psi _{11}\label{30_d}
	\end{align}
\end{subequations}
where $ \mathbf{U\Sigma }_a\mathbf{U}^H $ is the spectral decomposition of $ \breve{\mathbf{a}}_{r,t}\left( \theta ;r,\theta \right) \breve{\mathbf{a}}_{r,t}^{H}\left( \theta ;r,\theta \right)  $, $ \mathbf{UU}^H=\mathbf{I} $, and $ \mathbf{U} $ is the characteristic matrix.
Note that $ \mathbf{\Sigma }_a $ can be denoted as $ \mathbf{\Sigma }_a=\mathrm{diag}\left( \sigma _1,0,\cdots ,0 \right)  $ due to it is rank one. $ \sigma _1=\left\| \breve{\mathbf{a}}_{r,t}\left( \theta ;r,\theta \right) \right\| _{2}^{2} $.
From \eqref{30_c} to \eqref{30_d}, we use the property of diagonal matrix. $ \mathbf{\Psi }=\mathbf{U}^H\mathbf{R}_{zz}^{-1}\mathbf{U} $, and $ \Psi _{ii},i=1,\cdots ,MN $, is the diagonal elements of $ \mathbf{\Psi } $.

\begin{figure*}[htbp]
	\centering
	\subfigure {\includegraphics[width=0.88\textwidth]{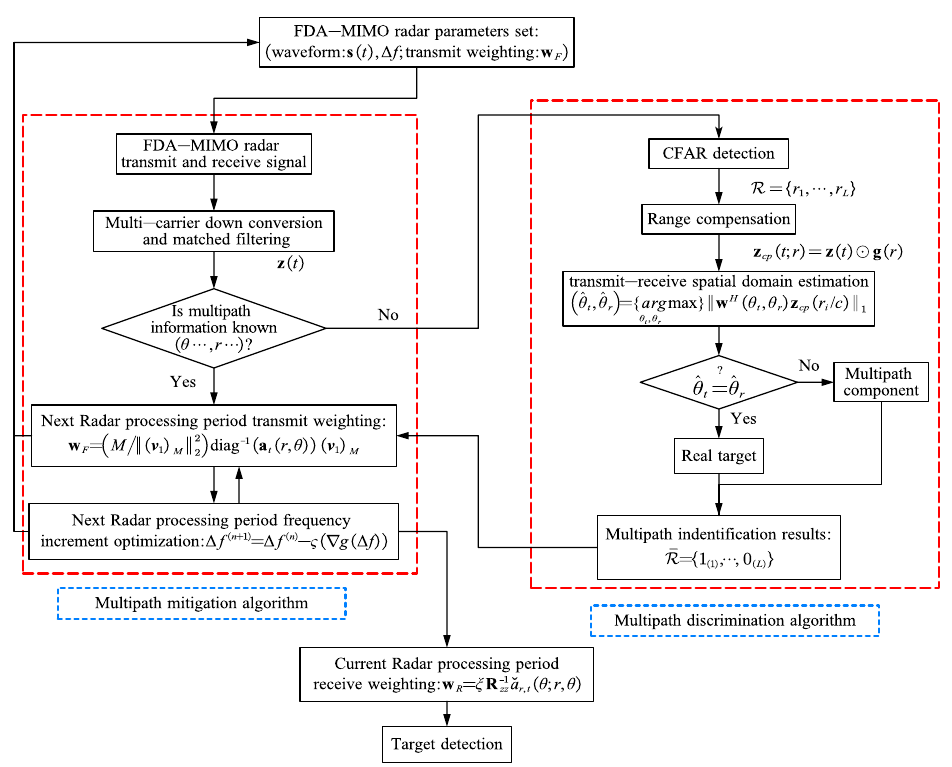}}
	\caption{Comprehensive signal processing diagram for the proposed algorithm based on FDA-MIMO radar.}
	\label{f.3}
\end{figure*}

Then, substituting \eqref{30_0} into \eqref{30_c}, we obtain
\begin{equation}\label{33}
	\mathbf{\Psi }=\mathbf{U}^H\mathbf{V}^H\mathbf{\Sigma }_{R}^{-1}\mathbf{VU}\\
	=\sum_{k=1}^{MN}{}\varrho _{k}^{-1}\mathbf{b}_k\mathbf{b}_{k}^{H}
\end{equation}
where $\left[ \mathbf{b}_1,\cdots ,\mathbf{b}_{MN} \right]= \left( \mathbf{VU} \right) ^H $ are the column split of the Hermite matrix $ \left( \mathbf{VU} \right) ^H $. Note that \eqref{33} is the rank one decomposition of $ 	\mathbf{\Psi } $, then substituting \eqref{33} into \eqref{30_d}, the original optimization \eqref{31} can be transformed as
\begin{equation}\label{34}
	\begin{cases}
		\underset{\Delta f}{\max}\,\,\sum_{k=1}^{MN}{}\varrho _{k}^{-1}\\
		s.t. \quad\mathbf{R}_{zz}^{}=\mathbf{V\Sigma }_R\mathbf{V}^H\\
		\qquad	\mathbf{\Sigma }_R=\mathrm{diag}\left( \varrho _1,\cdots ,\varrho _{MN} \right)\\
		\,\,   MN\sigma _{n}^{2}\leqslant \sum_{k=1}^{MN}{}\varrho _{k}^{}\leqslant MN\left[ \left| \eta _0 \right|^2\left( 1+\tilde{\rho}_0 \right) ^4+\sigma _{n}^{2} \right]\\
	\end{cases}
\end{equation}
The last constraint can be obtained by taking $ \ell _2 $ norm for the received signal \eqref{15}.
It is worth noting that the optimization objective in \eqref{34} is to maximize the trace of the inverse matrix $ \mathbf{R}_{zz}^{-1} $, and the constraint condition is the trace of the matrix $ \mathbf{R}_{zz} $. For a certain radar system, the trace of the received covariance matrix (also the energy of the received signal) is a constant (for a short period of time) under the stationary channel hypothesis. In other words, the energy of the received signal is only related to the transmit power, target scattering coefficient and channel attenuation. Therefore, the optimization problem of \eqref{34} can be partially equivalent to
\begin{equation}\label{35}
	\underset{\Delta f}{\min}\,\,tr\left( \mathbf{R}_{zz}^{} \right) 
\end{equation}
Using \eqref{15} and \eqref{24}, \eqref{35} can be transformed to (the detailed derivation process is shown in Appendix \ref{a2}),
\begin{equation}\label{36}
	\begin{split}
		\underset{\Delta f}{\min}\,\,\,\,g\left( \Delta f \right) =\left\| \breve{\mathbf{a}}_{t}^{H}\left( \bar{r},\theta ;\Delta f \right) \mathbf{R}_{ss}^{*}\left( 0;\Delta f \right) \right\| _{2}^{2}
		\\
		+\left\| \breve{\mathbf{a}}_{t}^{H}\left( \bar{r},\theta _s;\Delta f \right) \mathbf{R}_{ss}^{*}\left( 0;\Delta f \right) \right\| _{2}^{2}
		\\
		+\left| \tilde{\rho}_0 \right|^2\left\| \breve{\mathbf{a}}_{t}^{H}\left( r_s,\theta _s;\Delta f \right) \mathbf{R}_{ss}^{*}\left( 0;\Delta f \right) \right\| _{2}^{2}+\frac{\mathrm{tr}\left( \mathbf{R}_{ss}\left( 0 \right) \right)}{\left| \eta _0 \right|^2\left| \tilde{\rho}_0 \right|^2}
	\end{split}
\end{equation}
In other words, \eqref{36} also means finding a frequency increment that minimizes the projection of the equivalent transmit SV in the space corresponding to the several larger eigenvalues of the ambiguity matrix $ \mathbf{R}_{ss} $. Note that \eqref{36} is derivable, thus optimal $ \Delta f $ (assuming the frequency increment is within the specified range, usually $ \Delta f\in \left( 0,B_s \right)  $, where $ B_s  $ is the average bandwidth of the baseband signal) can be obtained by gradient descent algorithms\cite{Veatch2021linear}, i.e.,
\begin{equation}\label{37}
	\Delta f^{\left( n+1 \right)}=\Delta f^{\left( n \right)}-\varsigma \left( \nabla g\left( \Delta f \right) \right) 
\end{equation}
where $ \nabla g\left( \Delta f \right)  $ is the gradient of $ g\left( \Delta f \right)  $, $ \varsigma $ is the step size. To summarize, the maltipath mitigation algorithm is shown in Algorithm \ref{alg2}. In order to enhance the clarity of the methodology employed herein, we illustrate a complete and systematic radar signal processing workflow in Fig.\ref{f.3} Note that transmit weighting and frequency increment optimization may go through a process of multiple alternative solutions\cite{Veatch2021linear}.
\begin{algorithm}[htb] 
	\caption{Proposed multipath mitigation algorithm for FDA-MIMO} 
	\label{alg2} 
	\begin{algorithmic}[1] 
		\Require
		\\
		Receive signal after matched filtering: $\mathbf{z}\left( t \right)  $; 
		\\
		Target and multipath direction: $ \left( \theta _0,r \right)  $ and $ \theta _1, \cdots   $;
		\Ensure
		
		transmit weighting $\mathbf{w}_F  $, receive weighting $ \mathbf{w}_R $ and frequency increment $\Delta f  $ for next radar processing period;
		\item using \eqref{24} to obtain $ \mathbf{R}_{zz} $ and \eqref{29} to obtain $ \mathbf{w}_F  $; 
		\item solving $ \eqref{36} $ to obtain $ \Delta f$ (by \eqref{37}); 
		\item using \eqref{26} to obtain receive weighting and perform spatial filtering.
\\
		
		\Return :$\mathbf{w}_F  $, $\mathbf{w}_R  $  and $ \Delta f   $. 
	\end{algorithmic}
\end{algorithm} 

\section{COMPLEXITY ANALYSIS}
\label{s4}
In this section, we briefly analyze the complexity of the proposed method. Suppose we only consider the number of multiplication operations (MOs) (the main part of complexity calculation) and the number of MOs required for several typical mathematical operations can be seen in \cite{Zhang2019accuracycomplexity}.
\begin{enumerate}[]
	\item Algorithm 1: Signal processing for the proposed multipath discrimination,
	Assuming that the sampling points for fast time are $ L $, and without considering the complexity of the matched filtering and CFAR detection procedure, then we have:
	\begin{itemize}
		\item Range compensation (using \eqref{18} and \eqref{19}) needs $ \mathcal{O} \left( MNL \right)  $ MOs.
		\item  Transmit-receive spatial domain estimation (using \eqref{21} and \eqref{22}) needs $\mathcal{O} \left( M^3N^3 \right)  $ MOs.
		\item Multipath discrimination needs $ \mathcal{O} \left( L \right)  $ MOs.
	\end{itemize}
	
	\item Algorithm 2: Signal processing for the proposed multipath mitigation and parameter optimization,
	\begin{itemize}
		\item Receive spatial domain filtering (using \eqref{26}) need $ \mathcal{O} \left( M^3N^3 \right)  $ MOs.
		\item Next radar period transmit weighting calculation require $  \mathcal{O} \left(M^3N^3 \right)   $ MOs.
		\item Next radar period transmit frequency increment optimization require $  \mathcal{O} \left(M^3N^3 \right)   $ MOs.
	\end{itemize}
	\item Hence, the total complexity combine proposed algorithm 1 and 2 are $ \mathcal{O} \left(M^3N^3 \right)  $.
\end{enumerate}
 \section{SIMULATION RESULTS}
 \label{s5}
In this section, we use numerical simulation to verify the effectiveness of the proposed method for target detection in multipath environment. 
Here are some general parameters used in the following simulation:
\begin{enumerate}
	\item FDA-MIMO radar consists of $ M = 10 $ and $ N=10 $ array elements respectively, operating at $ f_0=10 $ GHz;
	\item The inter-spacing is $ d_t=d_r=0.5 \lambda_0=15$ mm;
	\item The reflection coefficient is $ \tilde{\rho}_0=0.5e^{-\pi /2} $, SNR$ =-10 $ dB, and the number of snapshots is $ 3072 $.
	\item the signal bandwidth is $ B_s=40 $ MHz, pulse repetition interval (PRI) is $ T_p= 25\, \mu s$, pulse duration is $ T_s =5\, \mu s $;
\end{enumerate}
\subsection{Example 1: Single target Multipath discrimination and mitigation}
\label{Exa1}
In the first scenario, we consider a uniformly moving object that is located at $ (r,\theta)$=(2 km, 70\degree) , and the projection distance between array and 
reflector is $ h_a=20 $ m. Thus, the range and angle of the reflected path can be calculated using \eqref{1} and \eqref{2} respectively, i.e., $ r_s=2.014 $ km, $ \theta _s= $ 111\degree.
\begin{figure*}[htbp]
	\centering
	\subfigure[]{
		\includegraphics[width=0.47\textwidth]{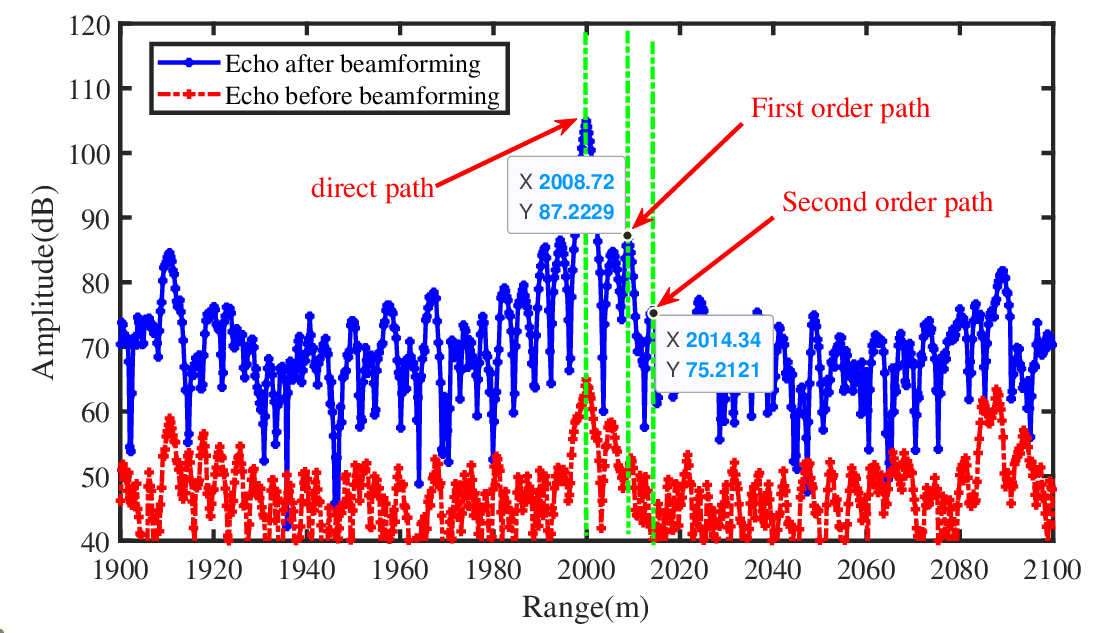}\label{f.5a}
	}
	\,
	\subfigure[]{
		\includegraphics[width=0.47\textwidth]{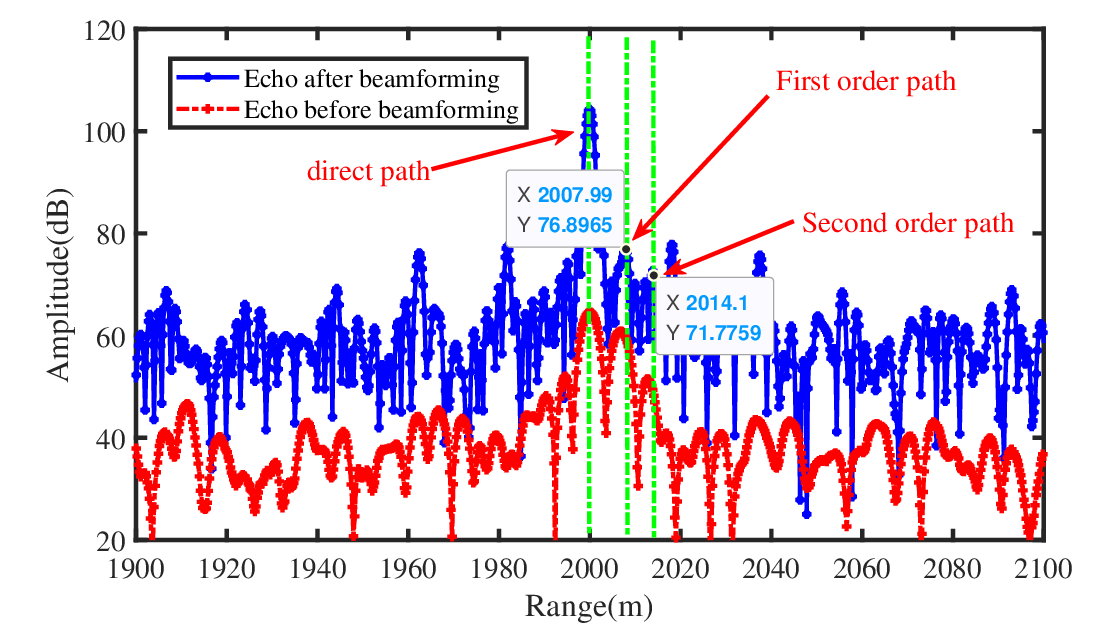}\label{f.5b}
	}
	\caption{Comparison of single channel matched filtering and receiving beamforming for two different transmit waveform radars. (a) MIMO radar using 11-bit Barker phase code with 256 sample waveforms, (b) FDA-MIMO radar using the same waveforms as Fig.\ref{f.5a} with frequency increment 20 MHz.}
	\label{f.5}
\end{figure*}
\subsubsection{verification of algorithm 1}
According to the identification process of algorithm \ref{alg1}, we compare the single-channel matched filtering and beamforming results of MIMO radar and FDA-MIMO radar in Fig.\ref{f.5}. In Fig.\ref{f.5a}, 11-bit Barker code\cite{Chen2023joint} is used as the orthogonal baseband waveform of MIMO radar, while in Fig.\ref{f.5b}, 11-bit Barker code with a certain frequency increment ($ \Delta f=0.49B_s=19.6 $ MHz) is used as the baseband waveform of FDA-MIMO radar. The red dotted line in Fig.\ref{f.5a} and \ref{f.5b} represent the results of the matched filtering of the received single channel echo. The blue solid line represents the result of receiving beamforming (weighted as the SV of the desired direction) for the $ MN $ virtual channel of the MIMO radar. From Fig.\ref{f.5a} and \ref{f.5b}, we can see that for MIMO and FDA-MIMO, although the echo signal energy at the desired target is enhanced by receiving beamforming, the first-order and second-order multipath components corresponding to the real target are also enhanced, which is reflected in the sidelobe peaks in Fig.\ref{f.5}. In addition, the distance and amplitude of the first-order multipath and the second-order multipath are shown in Fig.\ref{f.5}. It can be seen that even without the parameter optimization of algorithm 2, FDA-MIMO radar has better suppression results for multipath components than MIMO radar. Under the condition that the expected target gain is constant, the first order multipath peak is reduced by 10 dB, and the second order multipath peak is reduced by 5 dB. 

Table \ref{table1} shows some results of cell average constant false alarm rate (CA-CFAR) detection\cite{2022modern} after beamforming for MIMO and FDA-MIMO.
\begin{table*}[]
	\caption{MIMO and FDA-MIMO radar CA-CFAR results}
	\centering
	\begin{tabular}{l|lllllllll}
		\hline
		MIMO radar CFAR results (m)     & 1944.3 & 1962.3 & 1981.8 & \textbf{2000.1} & \textbf{2007.1} & {2018.1}  & 2037.3 & 2055.6 & 2090 \\ \hline
		FDA-MIMO radar CFAR results (m) & 1910.5 & 1957.7 & 1967.2 & \textbf{1999.9} & \textbf{2008.7} &  2020.1 & 2036.5 &{2057.3}& 2089.3  \\ \hline
	\end{tabular}
	\label{table1}
\end{table*}
Both MIMO and FDA-MIMO radars detect possible targets at direct path, first-order multipath (shown in bold in Table.\ref{table1}) and suppress the second-order multipath component. For MIMO radar, distance information and angle information are separated. Therefore, in each separate domain, it cannot distinguish between real targets and false targets formed by multipath interference, which reduces the accuracy of subsequent target detection and tracking. On the contrary, for FDA, due to the range-angle coupling characteristics caused by its frequency offset, we can identify multipath targets from the joint domain.
\begin{figure*}[htbp]
	\centering
	\subfigure[]{
		\includegraphics[width=0.45\textwidth]{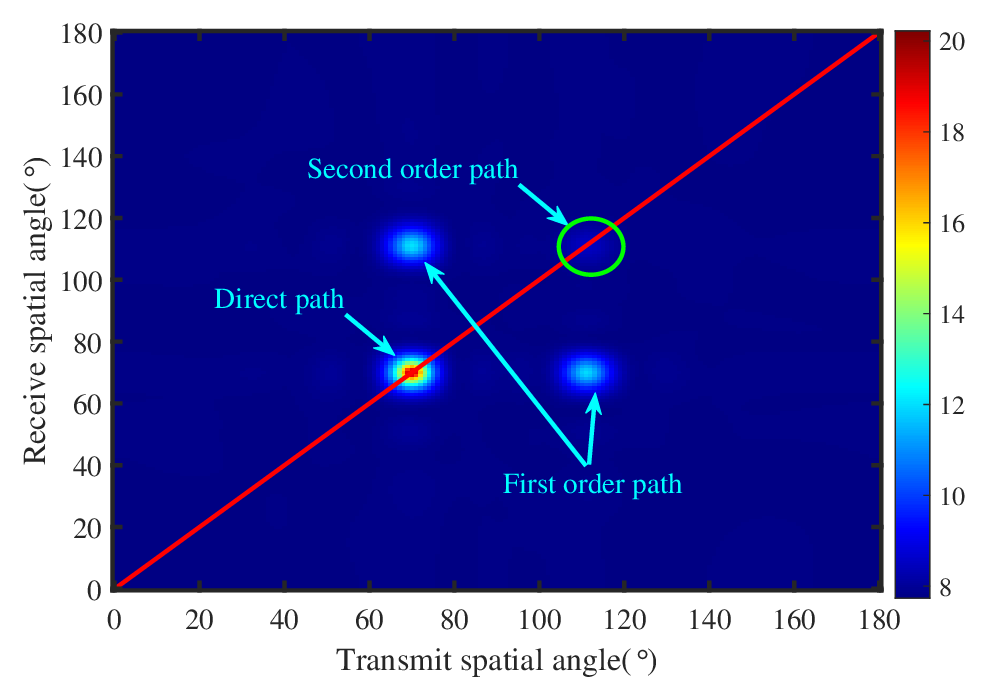}\label{f.6a}
	}
	\,
	\subfigure[]{
		\includegraphics[width=0.45\textwidth]{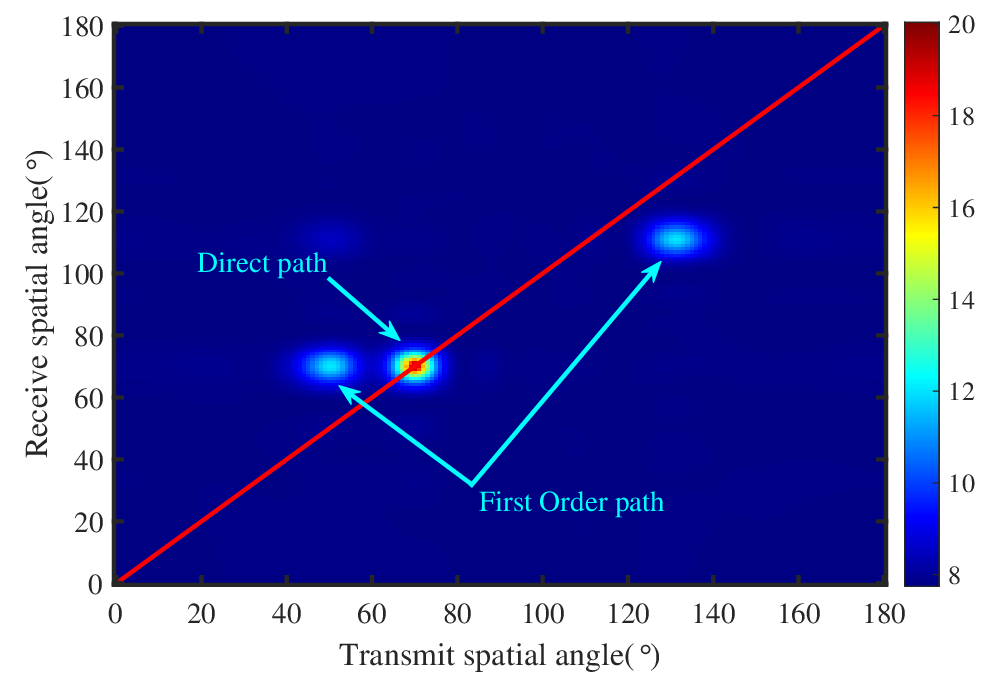}\label{f.6b}
	}
	\,
	\subfigure[]{
		\includegraphics[width=0.45\textwidth]{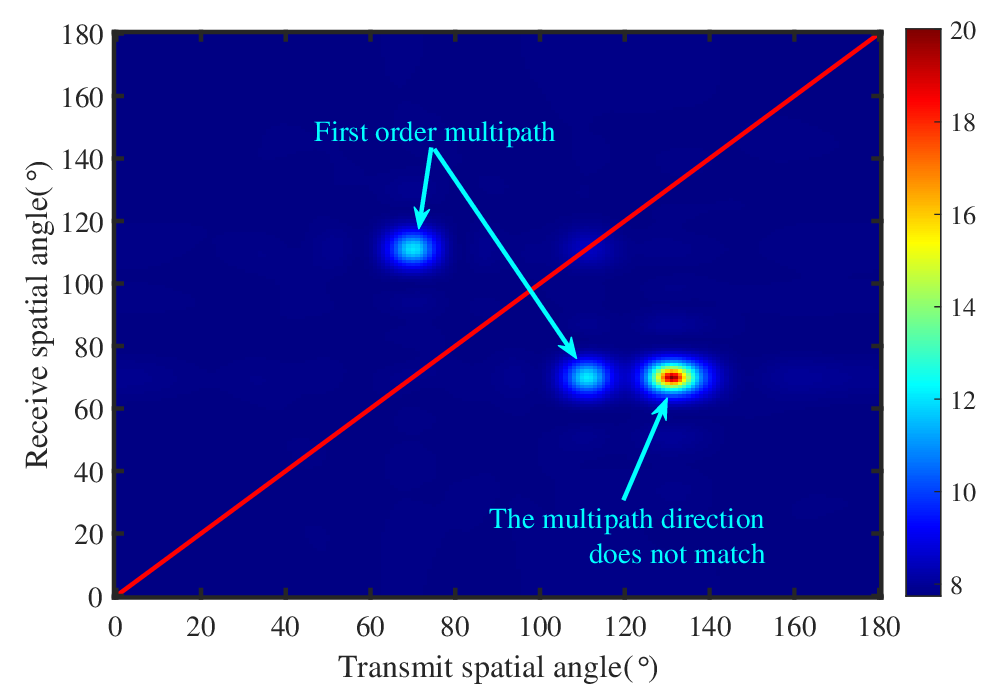}\label{f.6c}
	}
	\,
	\subfigure[]{
		\includegraphics[width=0.45\textwidth]{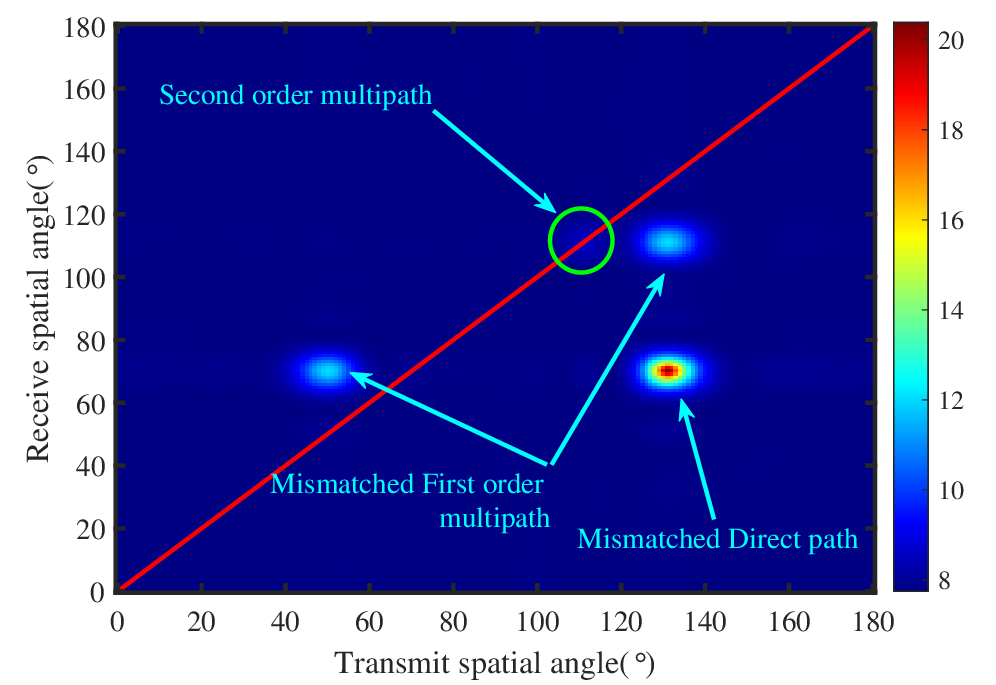}\label{f.6d}
	}
	\caption{Spatial spectrum of MIMO and FDA-MIMO radar (a) MIMO radar spatial spectrum without range information, (b) FDA-MIMO radar spatial spectrum with direct path range compensation $ 	\mathbf{g}(r) $, (c) is with first order multipath range compensation $ \mathbf{g}(\bar{r}) $, (d) is with second order multipath range compensation $ \mathbf{g}(r_s) $.}
	\label{f.6}
\end{figure*}
\begin{figure}[htbp]
	\centering
	\includegraphics[width=0.45\textwidth]{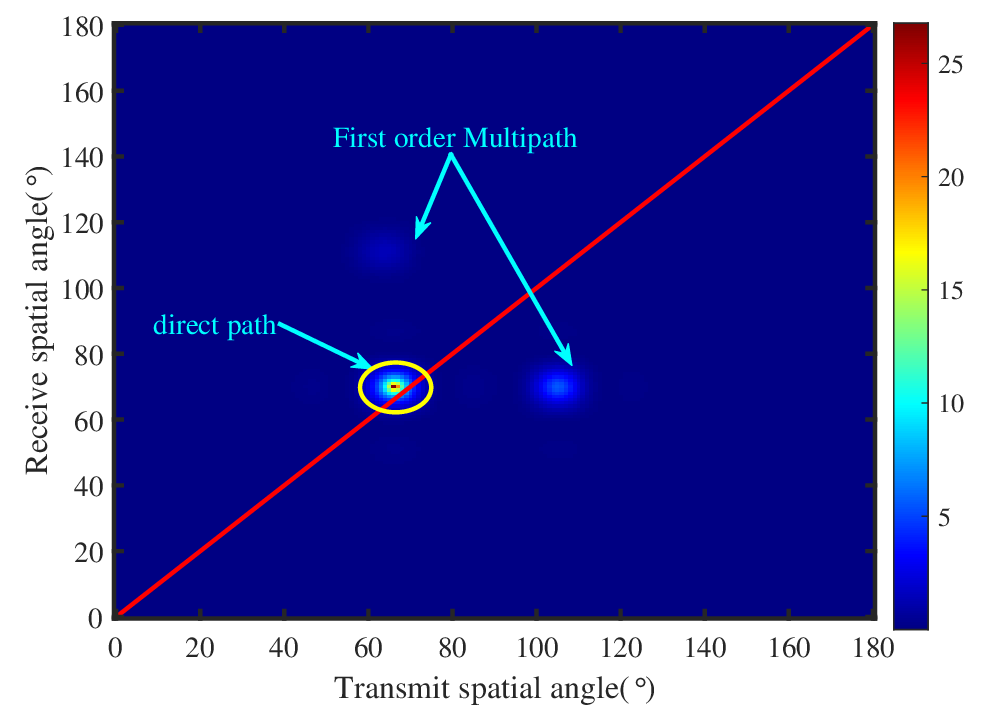}
	\caption{Spatial spectrum of FDA-MIMO radar using the method in \cite{Zheng2022signal}.}
	\label{f.7comp}
\end{figure}
Fig.\ref{f.6} shows the spatial angle information obtained by multi-channel spatial spectrum estimation of MIMO and FDA-MIMO radar. In Fig.\ref{f.6a}, there are strong target spatial spectral points at $ ( \hat{\theta}_t,\hat{\theta}_r ) =\left( 70\degree,70\degree \right)  $, and there are symmetric weak spatial spectral points at $ ( \hat{\theta}_t,\hat{\theta}_r ) =\left( 111\degree,70\degree \right) ,\left( 70\degree,111\degree \right)  $, which represent the two first-order multipath. Another, the spatial spectral points representing the weak second-order multipath signal appear at $ \left( 111\degree,111\degree \right)  $. Note that there is no distance compensation for MIMO radar, so it can only form one spatial spectrum.
Fig.\ref{f.6b}-\ref{f.6d} show the FDA-MIMO radar spatial spectrum estimation after different distance compensation. In Fig.\ref{f.6b}, after the distance compensation at the real target, its spatial spectrum peaks at the 70\degree\, diagonal, indicating that the current distance is the real target.
Another, in Fig.\ref{f.6c}, after using the first order multipath distance compensation, due to the distance difference between the first order multipath target and the real target, the first order multipath distance can not fully compensate for the distance-dependent SV corresponding to the real target, so that the real target deviates from the diagonal in Fig.\ref{f.6c}. For the multipath component, the transmitting angle and the receiving angle are different, so it can be determined that there is no real target at the current distance. In Fig.\ref{f.6d}, since the echo amplitude of the second order multipath component is much smaller than that of the real target and the first order multipath, the strongest amplitude point also deviates from the diagonal, which will not cause misjudgment.

In addition, in Fig.\ref{f.7comp}, we show the transmit-receive spatial spectrum obtained using the general music method in \cite{Zheng2022signal}. By comparing Fig.\ref{f.7comp} and Fig.\ref{f.5a} and \ref{f.5b}, it can be found that although the spatial angular resolution obtained by MUSIC spectrum is higher, due to the range-angle dependence of FDA, the peak value of its spatial spectrum deviates from the diagonal of the target, which may cause missed detection. The algorithm proposed in this paper has a smaller angle estimation error due to the distance compensation \cite{Lan2021singlesnapshot}. On the other hand, the MUSIC algorithm needs to estimate the noise subspace. In the case of multiple targets, it is difficult to determine the dimension of the noise subspace. However, our algorithm does not require subspace estimation and is more suitable for target detection and parameter estimation in multi-target scenarios (see Section \ref{Ex2}).
\subsubsection{verification of algorithm 2}
It is worth noting that algorithm \ref{alg1} proposed in this paper may fail in the case of a strong second-order multipath effect. Therefore, in order to further suppress the influence of second-order multipath on target detection, we propose algorithm \ref{alg2}, which reduces the energy radiation in the multipath direction by alternately optimizing the transmit waveform (frequency increment) and transmit weighting, so as to achieve the effect of multipath mitigation.
\begin{figure}[htbp]
	\centering
	\includegraphics[width=0.47\textwidth]{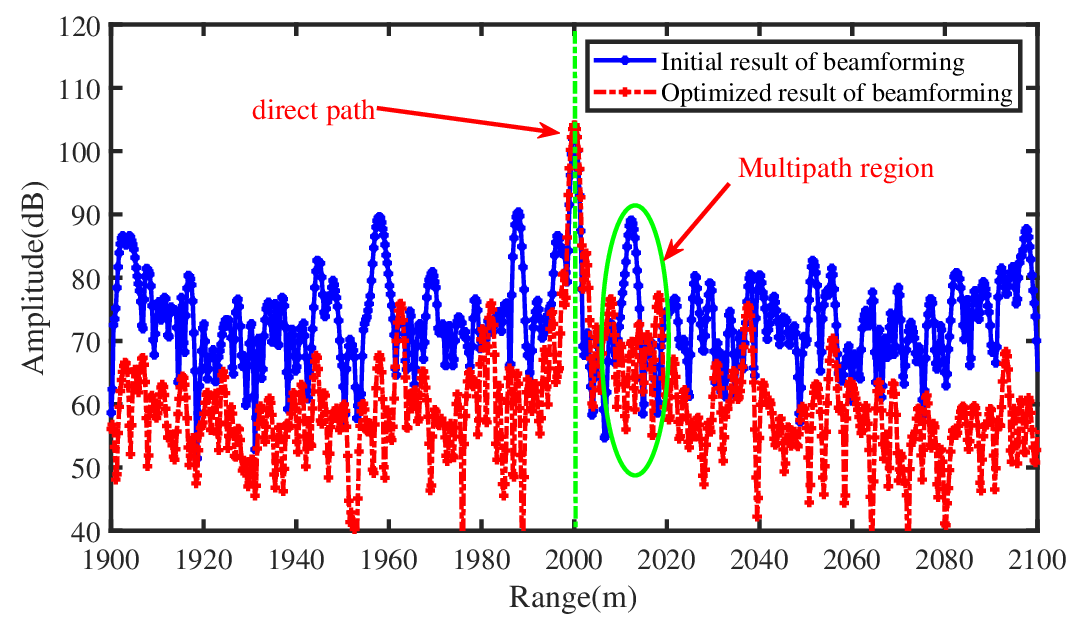}
	\caption{Receiving beamforming for FDA-MIMO radars with transmit weighting and frequency increment optimization using \eqref{29} and \eqref{36}.}
	\label{f.7}
\end{figure}
In Fig.\ref{f.7}, the blue solid line represents the initial beamforming results with uniform weighting and no frequency increment optimization (initially $ \Delta f=0.8B_s=32 $ MHz, and different initial frequencies have little effect on the final results), while the red dotted line represents the beamforming results after optimization (optimal $ \Delta f=16.4 $ MHz). As can be seen in Fig.\ref{f.7}, compared to the initial parameter waveform, the sidelobe level of the one-dimensional range profile is on average reduced by about 10 dB after weighting and parameter optimization, while maintaining the same gain at the target location. In addition, the amplitudes of the first and second order multipath components are reduced by about 15 dB. These simulation results demonstrate the effectiveness of the proposed algorithm.
\subsection{Example 2: Multiple target Multipath discrimination}
\label{Ex2}
\begin{figure}[htbp]
	\centering
	\includegraphics[width=0.47\textwidth]{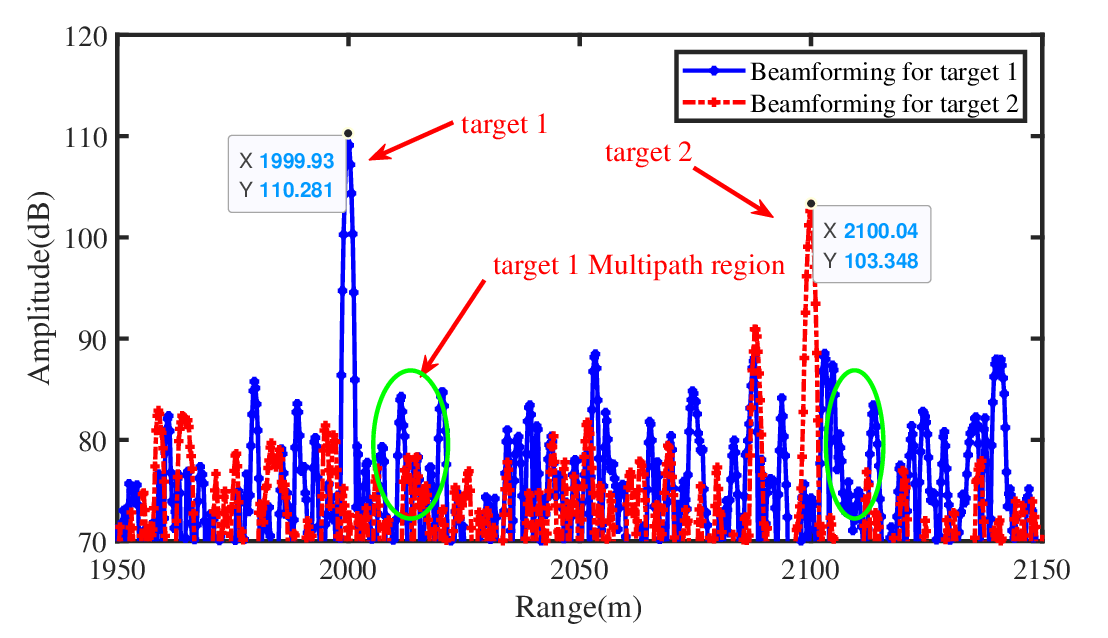}
	\caption{Receiving beamforming for FDA-MIMO radars of two targets.}
	\label{f.8}
\end{figure}
\begin{figure*}[htbp]
	\centering
	\subfigure[]{
		\includegraphics[width=0.45\textwidth]{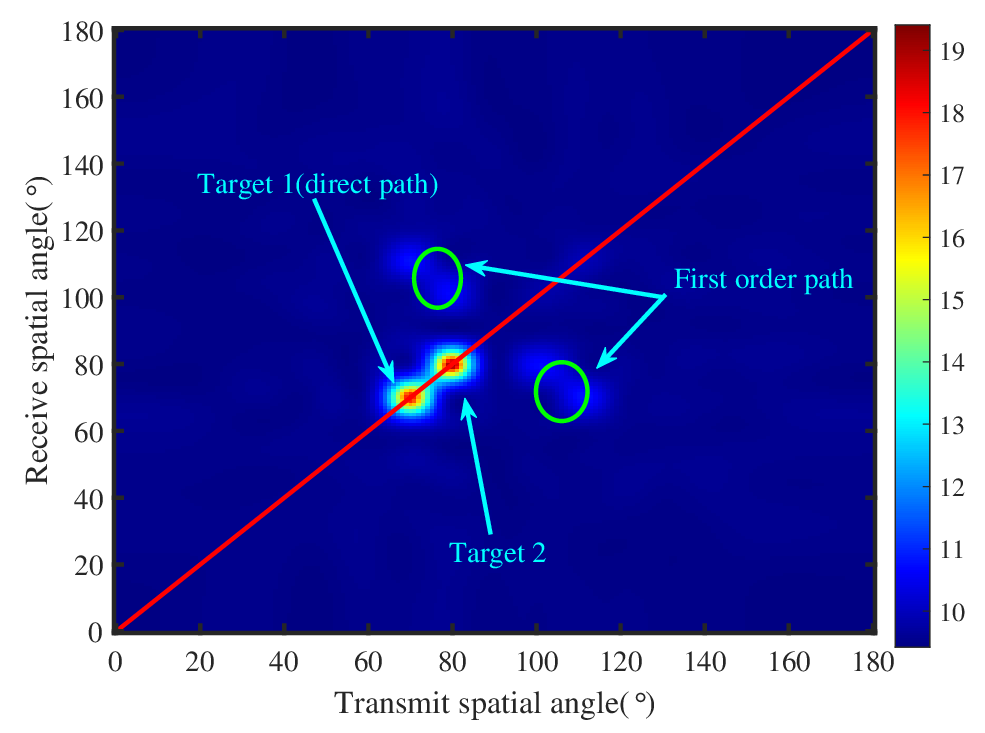}\label{f.9a}
	}
	\,
	\subfigure[]{
		\includegraphics[width=0.45\textwidth]{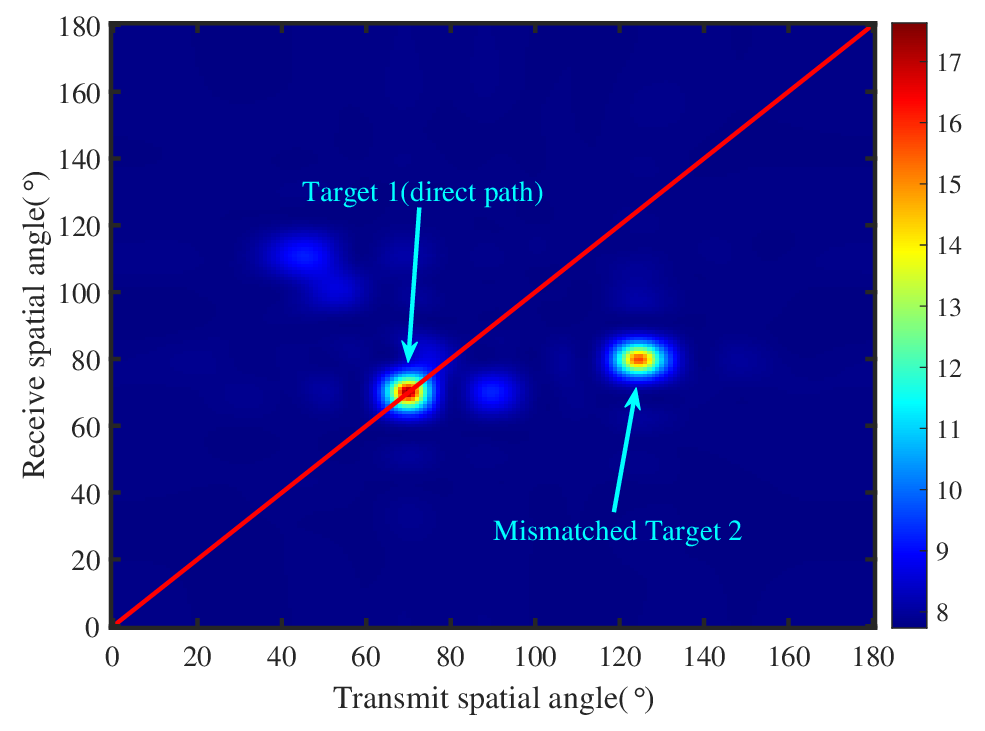}\label{f.9b}
	}
	\,
	\subfigure[]{
		\includegraphics[width=0.45\textwidth]{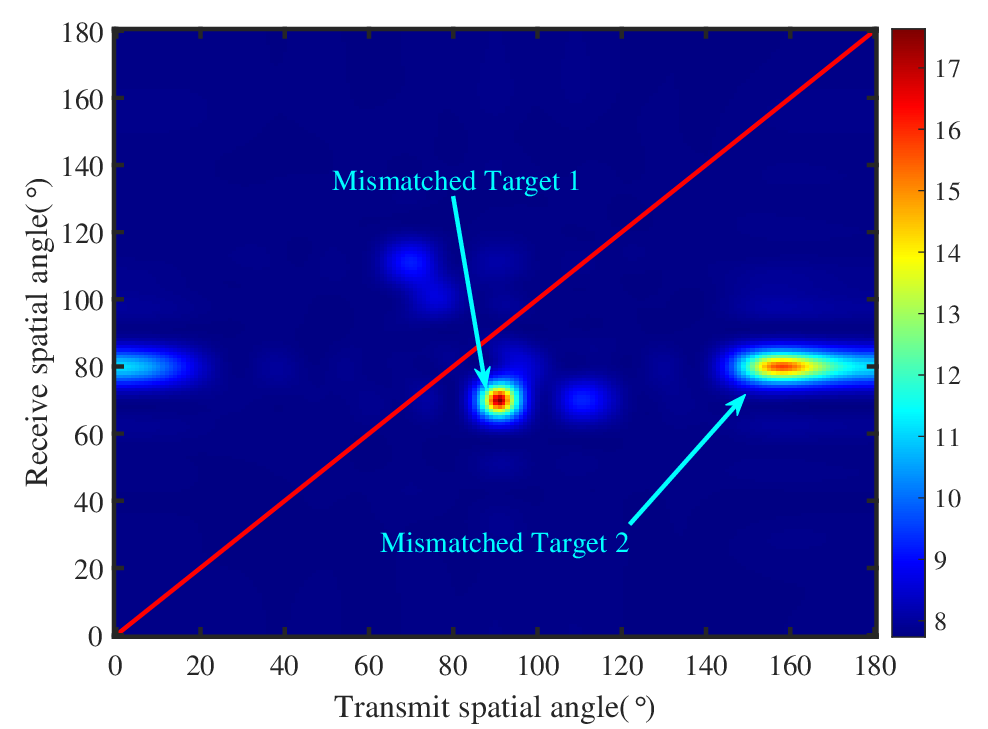}\label{f.9c}
	}
	\,
	\subfigure[]{
		\includegraphics[width=0.45\textwidth]{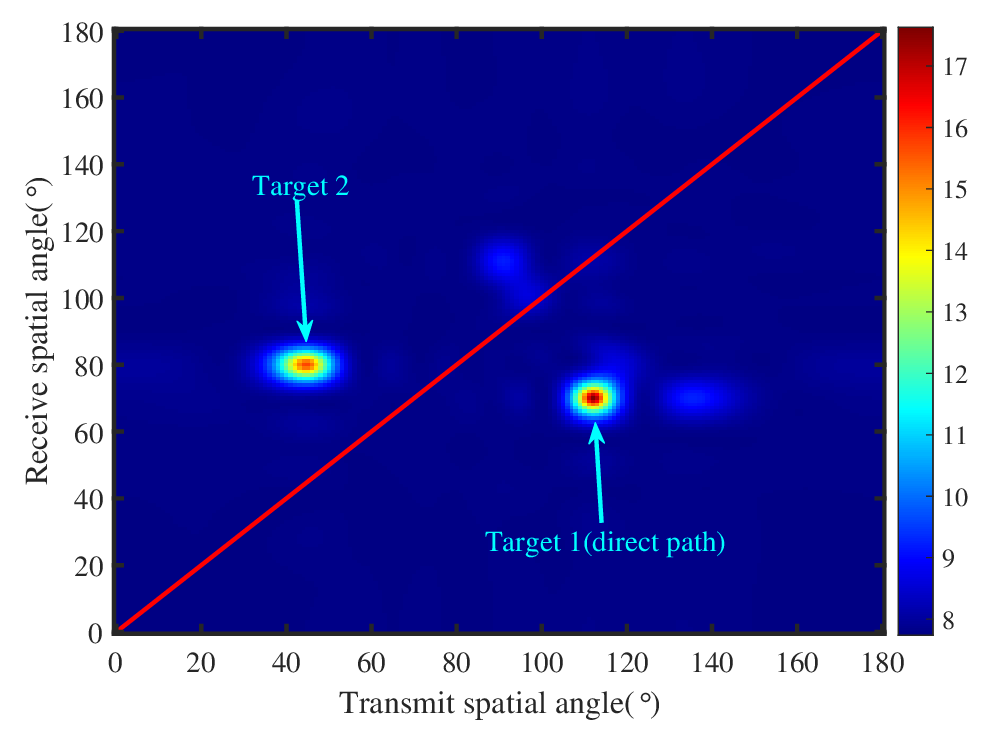}\label{f.9d}
	}
	\caption{Spatial spectrum of MIMO and FDA-MIMO radar with two targets (a) MIMO radar spatial spectrum without range information, (b) FDA-MIMO radar spatial spectrum with direct path range compensation $ 	\mathbf{g}(r) $ of target 1, (c) is with first order multipath range compensation $ \mathbf{g}(\bar{r}) $ of target 1, (d) is with second order multipath range compensation $ \mathbf{g}(r_s) $ of target 1.}
	\label{f.9}
\end{figure*}
In section \ref{Exa1}, we assumed that there was only one target in the scene. However, the effectiveness of the proposed algorithm for scenarios with multiple targets was unknown. Therefore, in this example, we consider a scenario with two targets at different angles and distances, together with the first and second order multipath components generated by each target. Assume that Target 1 and Target 2 are located at $ (r,\theta)$=(2 km, 70\degree) and (2.1 km, 80\degree), respectively. The FDA-MIMO radar baseband waveform and frequency increment are consistent with those in section \ref{Exa1}. Fig.\ref{f.8} illustrates the one-dimensional range profiles obtained by applying beamforming weights of steering vectors corresponding to two target locations. Target 2 is further away from Target 1, resulting in its amplitude being approximately 7 dB lower as shown in Fig.\ref{f.8}. Notably, spatial filtering was not able to completely eliminate multipath components of targets, as evidenced by amplitude peaks of multipath components near the target (indicated by green ellipses in Fig.\ref{f.8}).

Fig. \ref{f.9} shows the two-target spatial spectrum of MIMO and FDA-MIMO. Fig.\ref{f.9a} displays the transmit and receive spatial spectrum under the MIMO waveform, where amplitude peaks of the spatial spectrum occur at the respective angles of the two targets. However, it fails to provide coupled distance and angle information. For simplicity, we only show the spatial spectrum compensated by the target 1 distance in Fig.\ref{f.9}, and omit the spatial spectrum compensated by the target 2 distance (both principles are consistent). Fig.\ref{f.9b}-\ref{f.9d} respectively show the FDA-MIMO transmit and receive spatial angles after applying distance compensation corresponding to target 1, and the first and second order multipath range of target 1. It is observed that in multi-target scenarios, the proposed algorithm can still accurately associate targets with their true angles, providing discriminative information for multipath interference. From another perspective of Fig.\ref{f.9d}, it can be seen that the proposed algorithm can still distinguish whether the current range cell has a real target from angle information when the CFAR detection false alarm occurs.
\section{CONCLUSION}
\label{s6}
This paper tackles the issue of multipath interference detection via the transformation into spatial spectrum estimation in the identified range cell, leveraging the range-angle covariance features of FDA-MIMO radar. This approach rectifies each received snapshot using the targeted distance SV. The rectified data is employed to calculate the covariance and ascertain the corresponding spatial spectrum. Any strong amplitude points with identical transmitting and receiving angles in the current spectral plot imply presence of a physical target in the targeted distance unit. Moreover, the echo intensity of the multipath component is decreased through jointly optimizing the array transmission weighting and waveform frequency increment, minimizing false alarms. The efficacy of this algorithm is verified via numerical simulation. Future studies will delve into multipath mitigation in urban or indoor settings.

\appendices
\section*{APPENDIX}
\subsection{derivation of \eqref{29}}
\label{a1}
Using \eqref{30_0}, the inverse matrix of $\mathbf{R}_{zz}  $ can be factorized as 
\begin{equation}\label{ap1}
	\mathbf{R}_{zz}^{-1}=\mathbf{V}^H\mathbf{\Sigma }_{R}^{-1}\mathbf{V}
\end{equation}
Then \eqref{28} can be transformed as
\begin{equation}\label{ap2}
	\begin{cases}
		\underset{\mathbf{w}_F,\Delta f}{\max}\,\,\,\psi \left( \mathbf{w}_F,\Delta f \right) \,=\sum_{k=1}^{MN}{}\varrho _{k}^{-1}\breve{\mathbf{a}}_{r,t}^{H}\boldsymbol{\nu }_k\boldsymbol{\nu }_{k}^{H}\breve{\mathbf{a}}_{r,t}\\
		s.t.\qquad \left\| \mathbf{w}_F \right\| _2=\sqrt{M}\\
	\end{cases}
\end{equation}
For simplicity, we omit the dependency term that $ \breve{\mathbf{a}}_{r,t}=\breve{\mathbf{a}}_{r,t}\left( \theta ;r,\theta \right)  $.
It is noted that our received covariance matrix contains part of the desired signal. Therefore, intuitively, the optimization goal of \eqref{ap2} should be corrected to make the equivalent transmit SV perpendicular to the SV space corresponding to multipath interference (rather than perpendicular to the desired space and multipath interference space). Then, since the eigenvalues corresponding to the multipath interference term in $ \mathbf{R}_{zz} $ are larger, and the eigenvalues corresponding to the noise in $ \mathbf{R}_{zz} $ are smaller, the weighting is the largest in the inverse matrix $ \mathbf{R}_{zz}^{-1} $. Therefore, as long as the equivalent transmit SV $  \breve{\mathbf{a}}_{r,t}$ is perpendicular to the feature space corresponding to multipath interference, \eqref{ap2} can be maximized. Therefore, we only need to make the emission equivalent SV parallel to the eigenvector corresponding to the maximum eigenvalue of $ \mathbf{R}_{zz}  $ to obtain the solution to the optimization problem, i.e.,
\begin{equation}\label{key}
	\breve{\mathbf{a}}_{r,t}\left( \theta ;r,\theta \right) =\mathbf{a}_r\left( \theta \right) \otimes \left[ \mathbf{w}_F\odot \mathbf{a}_t\left( r,\theta \right) \right] =\kappa \left( \boldsymbol{\nu }_1 \right) _M
\end{equation}
where $ \kappa  $ is the unknown coefficient to be solved. Another, taking into account the unit irradiation of the target orientation, $ \mathbf{w}_{F}^{H}\mathbf{a}_t\left( r,\theta \right) =M $.  Thus, the optimal transmit weighting can be regarded as the projection of the target direction SV in the feature space corresponding to the maximum eigenvalue of $ \mathbf{R}_{zz}  $, i.e.,
\begin{equation}\label{key}
	\mathbf{w}_F=\mathbf{a}_t\left( r,\theta \right) \left( \boldsymbol{\nu }_1 \right) _M\left( \boldsymbol{\nu }_1 \right) _{M}^{H}
\end{equation}
and $ \left( \boldsymbol{\nu }_1 \right) _M $ is the eigenvector corresponding to the maximum eigenvalue of $ \mathbf{R}_{zz}^{} $ which takes the first $ M $ values. Then, considering the energy constraints $ \left\| \mathbf{w}_F \right\| _2=\sqrt{M} $, we obtain \eqref{29}.
\subsection{derivation of \eqref{36}}
\label{a2}
Using \eqref{15} and \eqref{24}, we obtain \eqref{42}. 
\begin{figure*}[htbp] 
	\centering
	\begin{equation}\label{42}
		\begin{split}
			\mathbf{z}^H\left( t;\Delta f,\mathbf{w}_{T}^{} \right) \mathbf{z}\left( t;\Delta f,\mathbf{w}_{T}^{} \right) =\left| \eta _0 \right|^2\left\{ \left[ \mathbf{a}_r\left( \theta \right) \mathbf{a}_{r}^{H}\left( \theta \right) \right] \otimes \left[ \mathbf{R}_{ss}^{T}\left( t-\tau _0 \right) \breve{\mathbf{a}}_t\left( r,\theta \right) \breve{\mathbf{a}}_{t}^{H}\left( r,\theta \right) \mathbf{R}_{ss}^{*}\left( t-\tau _0 \right) \right] \right. 
			\\
			+\left| \tilde{\rho}_0 \right|^2\left[ \mathbf{a}_r\left( \theta _s \right) \mathbf{a}_{r}^{H}\left( \theta _s \right) \right] \otimes \left[ \mathbf{R}_{ss}^{T}\left( t-\tau _1 \right) \breve{\mathbf{a}}_t\left( \bar{r},\theta \right) \breve{\mathbf{a}}_{t}^{H}\left( \bar{r},\theta \right) \mathbf{R}_{ss}^{*}\left( t-\tau _1 \right) \right] 
			\\
			+\left| \tilde{\rho}_0 \right|^2\left[ \mathbf{a}_r\left( \theta \right) \mathbf{a}_{r}^{H}\left( \theta \right) \right] \otimes \left[ \mathbf{R}_{ss}^{T}\left( t-\tau _1 \right) \breve{\mathbf{a}}_t\left( \bar{r},\theta _s \right) \breve{\mathbf{a}}_{t}^{H}\left( \bar{r},\theta _s \right) \mathbf{R}_{ss}^{*}\left( t-\tau _1 \right) \right] 
			\\
			+\left. \left| \tilde{\rho}_0 \right|^4\left[ \mathbf{a}_r\left( \theta _s \right) \mathbf{a}_{r}^{H}\left( \theta _s \right) \right] \otimes \left[ \mathbf{R}_{ss}^{T}\left( t-\tau _2 \right) \breve{\mathbf{a}}_t\left( r_s,\theta _s \right) \breve{\mathbf{a}}_{t}^{H}\left( r_s,\theta _s \right) \mathbf{R}_{ss}^{*}\left( t-\tau _2 \right) \right] \right\} +\left( \mathbf{1}_N\mathbf{1}_{N}^{T} \right) \otimes \mathbf{R}_{ss}\left( 0 \right) 
		\end{split}
	\end{equation}
\end{figure*}
Here $ \tau _1=2\bar{r}/c,\tau _2=2r_s/c $, we mainly consider the autocorrelation of the SV of the target and the multipath, while ignoring the cross-correlation between them.
Note that we mainly consider the test cell at the location of the target or multipath component, thus, substituting $  t=\tau _0,\tau _1,\tau _2$ into \eqref{42}, respectively. We obtain
\begin{equation}\label{43}
	\begin{split}
		&\,\,\mathrm{tr}\left( \mathbf{R}_{zz}^{} \right) \approx \left| \eta _0 \right|^2N\left\{ \left\| \breve{\mathbf{a}}_{t}^{H}\left( r,\theta \right) \mathbf{R}_{ss}^{*}\left( 0 \right) \right\| _{2}^{2} \right. 
		\\
		&\qquad\qquad+\left| \tilde{\rho}_0 \right|^2\left\| \breve{\mathbf{a}}_{t}^{H}\left( \bar{r},\theta \right) \mathbf{R}_{ss}^{*}\left( 0 \right) \right\| _{2}^{2}
		\\
		&\qquad\qquad+\left| \tilde{\rho}_0 \right|^2\left\| \breve{\mathbf{a}}_{t}^{H}\left( \bar{r},\theta _s \right) \mathbf{R}_{ss}^{*}\left( 0 \right) \right\| _{2}^{2}
		\\
		&+\left. \left| \tilde{\rho}_0 \right|^4\left\| \breve{\mathbf{a}}_{t}^{H}\left( r_s,\theta _s \right) \mathbf{R}_{ss}^{*}\left( 0 \right) \right\| _{2}^{2} \right\} +N\cdot \mathrm{tr}\left( \mathbf{R}_{ss}\left( 0 \right) \right) 
	\end{split}
\end{equation}
To obtain \eqref{43}, for each summation term in \eqref{43}, a similar derivation process is used as follows:
\begin{equation}\label{key}
	\begin{split}
		&	\mathrm{tr}\left( \begin{array}{c}
			\left[ \mathbf{a}_r\left( \theta \right) \mathbf{a}_{r}^{H}\left( \theta \right) \right] \otimes\\
			\left[ \mathbf{R}_{ss}^{T}\left( t-\tau _0 \right) \breve{\mathbf{a}}_t\left( r,\theta \right) \breve{\mathbf{a}}_{t}^{H}\left( r,\theta \right) \mathbf{R}_{ss}^{*}\left( t-\tau _0 \right) \right]\\
		\end{array} \right) 
		\\
		&	=N\mathrm{tr}\left( \left[ \mathbf{R}_{ss}^{T}\left( t-\tau _0 \right) \breve{\mathbf{a}}_t\left( r,\theta \right) \breve{\mathbf{a}}_{t}^{H}\left( r,\theta \right) \mathbf{R}_{ss}^{*}\left( t-\tau _0 \right) \right] \right) 
		\\
		&	=N\mathrm{tr}\left( \left[ \breve{\mathbf{a}}_{t}^{H}\left( r,\theta \right) \mathbf{R}_{ss}^{*}\left( t-\tau _0 \right) \mathbf{R}_{ss}^{T}\left( t-\tau _0 \right) \breve{\mathbf{a}}_t\left( r,\theta \right) \right] \right) 
		\\
		&	=N\left\| \breve{\mathbf{a}}_{t}^{H}\left( r,\theta \right) \mathbf{R}_{ss}^{*}\left(  t-\tau _0 \right) \right\| _{2}^{2}
	\end{split}
\end{equation}
In order to ensure the full illumination of the target area, the optimization objective function of \eqref{35} should be modified to remove the target energy term $ \left\| \breve{\mathbf{a}}_{t}^{H}\left( r,\theta \right) \mathbf{R}_{ss}^{*}\left( 0 \right) \right\| _{2}^{2} $. Therefore, the final optimization problem can be expressed as \eqref{36}.

\bibliographystyle{ieeetaes}
\bibliography{Ref}




\end{document}